\documentclass[aps,12pt, showkeys]{revtex4}  
\usepackage{graphicx}  
\usepackage{amsmath,amsthm,amsfonts,amscd}  
  
\renewcommand{\vec}[1]{\mbox{\boldmath$#1$}}

\newcommand{\sref}[1]{section~\ref{#1}}
\newcommand{\fref}[1]{figure~\ref{#1}}

\newcommand{\Fref}[1]{Figure~\ref{#1}}

\usepackage{float}  
\begin{document}
\vspace{20mm}   
\title{Trapped Particle Stability for the Kinetic Stabilizer}   
\author{H. L. Berk$^1$ and J. Pratt$^2$}   
\affiliation{$^1$ Institute for Fusion Studies, The University of Texas at Austin, Austin, TX  78639, USA  }  
\affiliation{$^2$ Max-Planck-Institut f\"ur Plasmaphysik, 85748 Garching, Germany  and Max-Planck-Institut 
f\"ur Sonnensystemforschung, 37191 Katlenburg-Lindau, Germany}
\email{Jane.Pratt@ipp.mpg.de}
\begin{abstract}
A kinetically stabilized axially symmetric tandem mirror (KSTM) uses the
momentum flux of low-energy, unconfined particles that sample only the 
outer end-regions of the mirror plugs, where large favorable field-line
curvature exists.  The window of operation is determined for achieving
MHD stability with tolerable energy drain from the kinetic stabilizer. 
Then MHD stable systems are analyzed for stability of the trapped
particle mode. This mode is characterized by the detachment of the
central-cell plasma from the kinetic stabilizer region without inducing
field-line bending. Stability of the trapped particle mode is sensitive
to the electron connection between the stabilizer and the end plug. It
is found that the stability condition for the trapped particle mode is
more constraining than the stability condition for the MHD mode, and it
is challenging to satisfy the required power constraint.  Furthermore a severe power drain may arise from the
necessary connection of low-energy electrons in the kinetic stabilizer to the central region.
\end{abstract}
%\pacs{52.55.-s, 28.52.-s}
\keywords{tandem mirrors, trapped particle mode, kinetically stabilized tandem mirror, MHD stability \\
submitted to Nuclear Fusion, March 2011}
\maketitle
\section{Introduction}

The tandem mirror magnetic-fusion confinement system is a
nearly-cylindrical solenoid terminated by a set of plug cells. These
plugs consist of simple axisymmetric mirror fields. The earliest
stability theories \cite{rosenlong} predicted that symmetric mirror
machines, including tandem mirrors would be MHD unstable; experiments in
the 1980s on both TARA \cite{posttara} and PHAEDRUS \cite{phaedrus}
\cite{Molvik} tandem mirror facilities confirmed that MHD instability
occurs. Recently, a new stability innovation, the kinetic stabilizer,
has been proposed to stabilize an axisymmetric tandem mirror. However,
the mechanism by which MHD stabilization is achieved in the kinetic
stabilizer may make the system susceptible to a rapidly growing trapped
particle instability\cite{prattdiss09}. Here we investigate this
possibility.

The kinetic stabilizer is a design proposed by R. F. Post
\cite{postinbook}, inspired by the work of D. Ryutov and experimental
evidence from the Gas Dynamic Trap (GDT), a single mirror experiment at
Novosibirsk \cite{ryutovgdt} \cite{ryutov1}. Ryutov established
theoretically that an otherwise MHD-unstable plasma confined between
symmetric mirrors can be stabilized if there exists sufficient momentum flux from the
effluent plasma on the expanding positive curvature field-lines outside
the mirrors. The momentum flux generalizes the role of the
pressure tensor that appears in standard MHD theory. This technique of
stabilization was experimentally confirmed using the axisymmetric GDT
\cite{ani97}. The GDT operates in a high-collisionality regime in order
to keep the loss-cone full. Ryutov noted that the effect of plasma
momentum flux flowing out of the ends of the machine in the
positive-curvature expanding-field region outside the mirrors is
sufficiently strong to overcome the destabilizing curvature contribution
from the central part of the plasma. This stabilization has been
experimentally demonstrated \cite{Ivanov}. It has been shown that MHD
stability persists to a moderately high-beta value $\beta \sim 0.3$ arising just from the exiting momentum flux.

In the GDT the loss-cone is filled by relatively strong collisions in
the central plasma region; with sufficient effluent flow, MHD stability
is established. However, in the typical tandem mirror the ends of the
machine are designed to confine a long mean-free-path plasma which only
produces a weak effluent. Thus there is a negligible momentum flux in
the expander region. As a result, the stabilization mechanism designed
for the GDT, \emph{i.e.} stabilization from the momentum flux arising
from the effluent plasma, does not simply arise in a tandem mirror of
conventional design. The kinetic-stabilizer concept proposes to solve
the problem of low effluence by using external ion-beams injected
axially into the machine. These kinetic-stabilizer ion-beams are
injected at small pitch-angles to the magnetic field so that the beam
can propagate towards the higher magnetic field and then reflect before
reaching the principal confinement region of the mirror plasma. The ion
beam then transits out of the machine.  While entering and exiting the
ends of the machine, the beam forms an unconfined plasma with an
enhanced momentum flux in a localized region of favorable curvature.
%FIG. 1
\begin{figure}[H]
\begin{center}
\includegraphics[scale=.8]{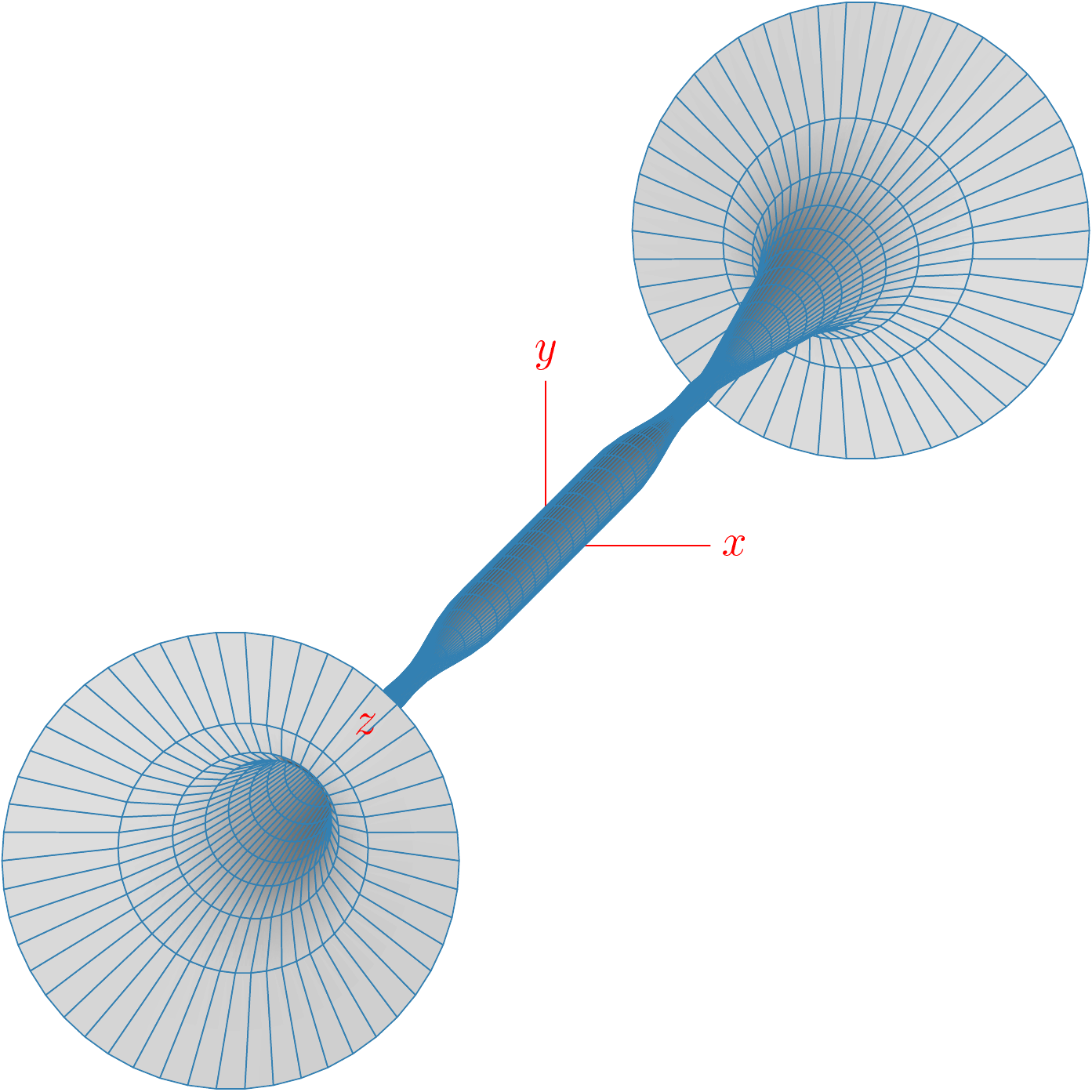}
\hspace{1mm}
\includegraphics[scale=.9]{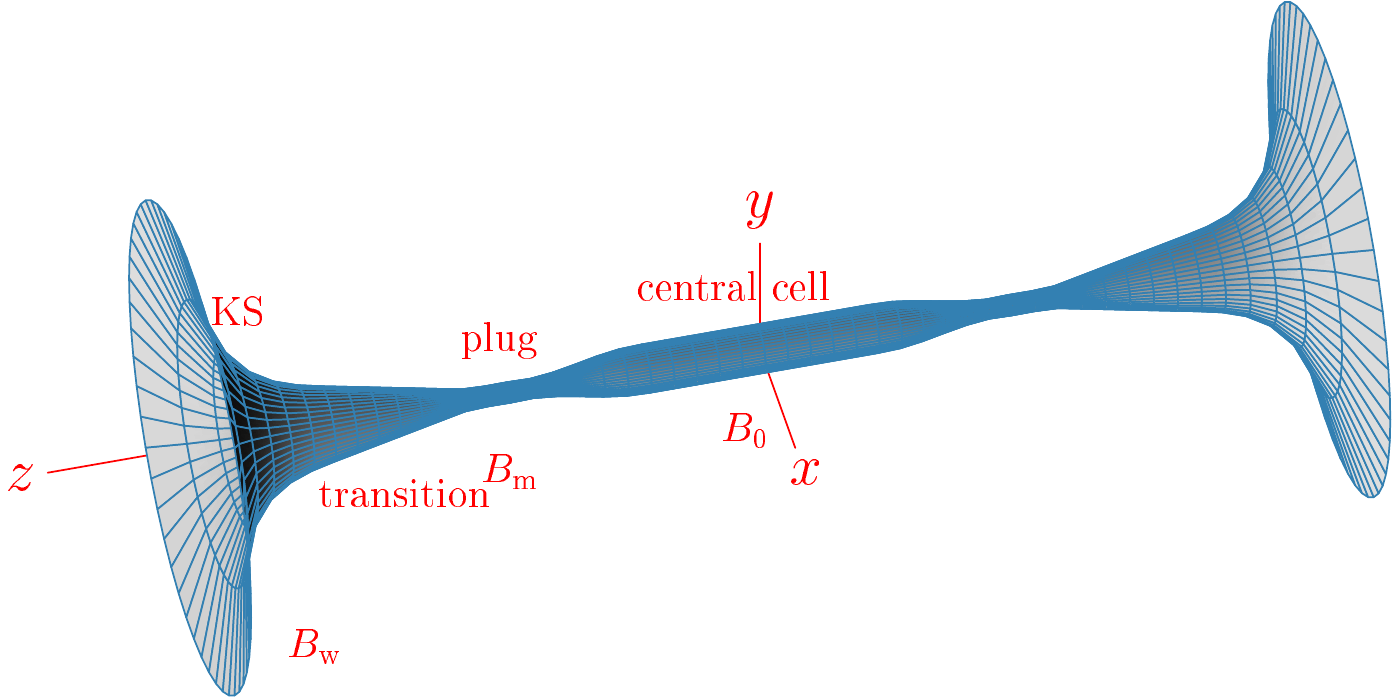}
\end{center}
\caption{Axisymmetric magnetic flux surfaces in the KSTM.
\label{post3d}}
\end{figure}

\begin{figure}[H]
\begin{center}
\includegraphics[scale=.8]{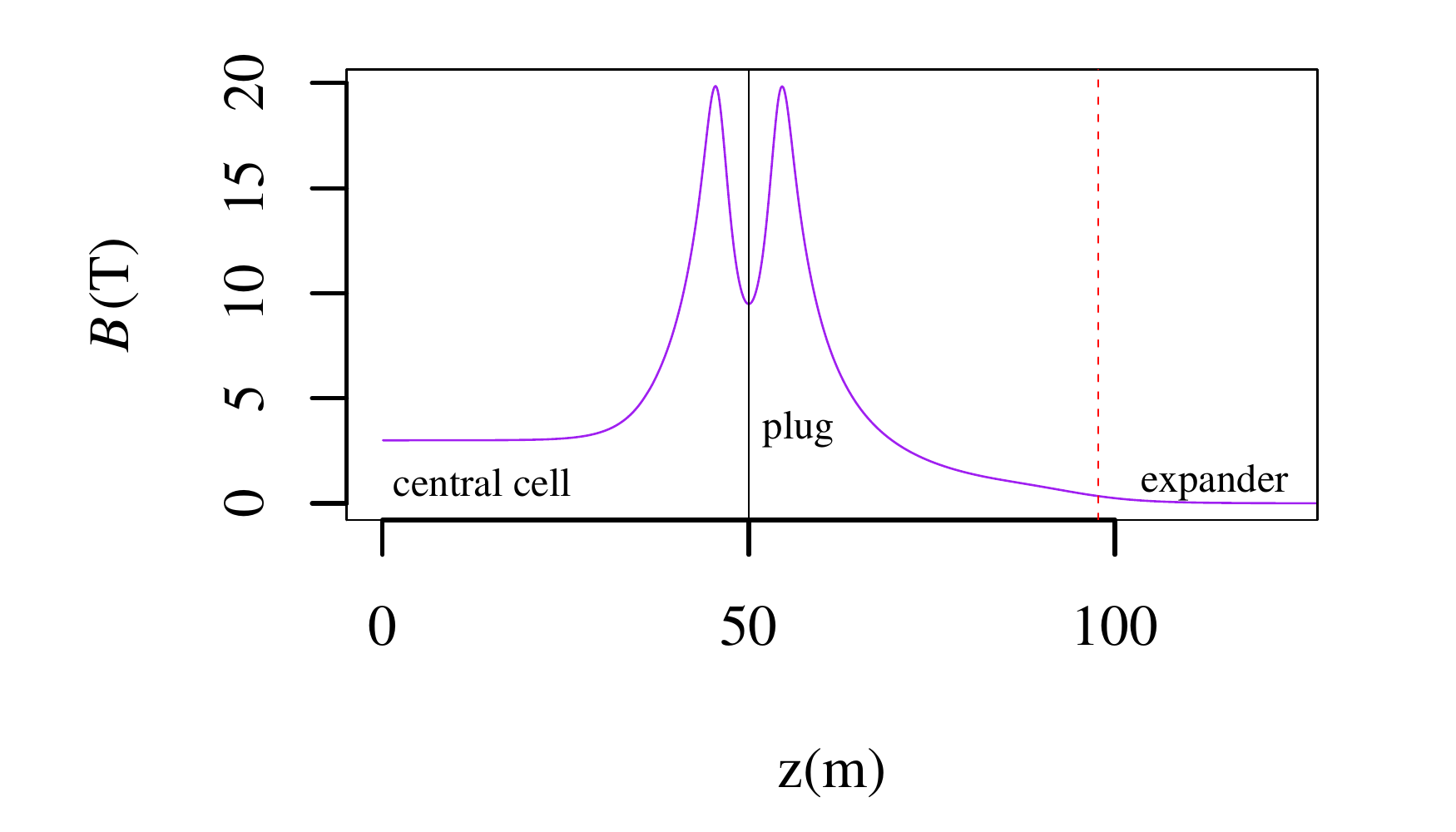}
\end{center}
\caption{Axial magnetic field $B$, in the right half of the proposed
KSTM reactor. The field-line curvature is negligible to the left of the
dashed line. The favorable curvature of the expander region lies to the
right of the dashed line.  A solid black line marks the center of the
mirror plug.}
\label{fkstm}
\end{figure}

\begin{figure}[H]
\includegraphics[scale=.95]{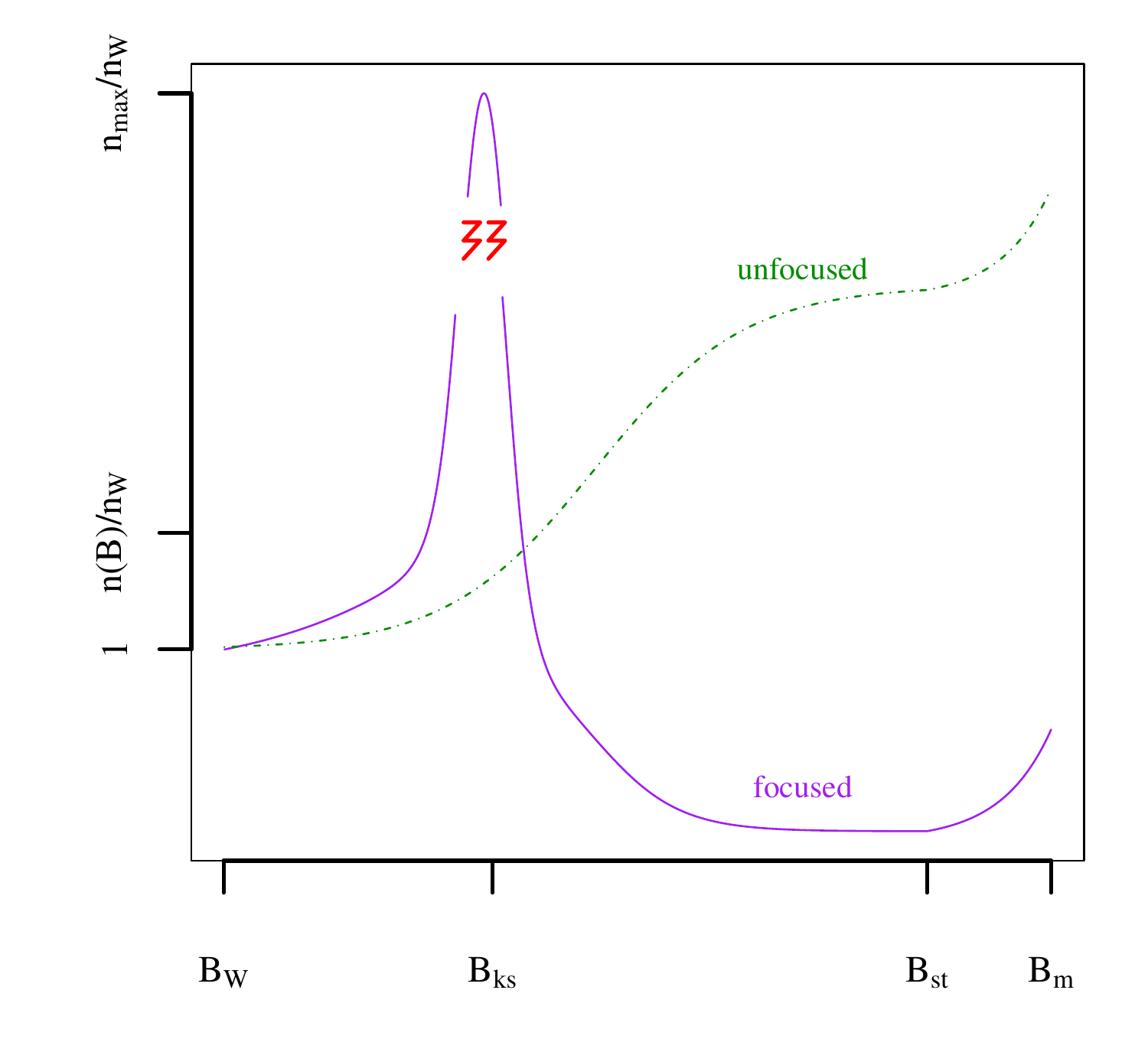}
\caption{A conceptual sketch of the density in the expander region of
the KSTM.  The shape of the density at the wall $B_{\mathrm{W}}$, the
entrance to the region of favorable curvature $B_{\mathrm{ks}}$,
plateau, stand-off point $B_{\mathrm{st}}$, and the maximum of the
plug $B_{\mathrm{m}}$ are shown.  }
\label{denff2}
\end{figure}

The proposed KSTM reactor is a simple axisymmetric tandem mirror. The
plasma in the plug regions possesses higher density and energy than the
central-cell plasma, producing ambipolar traps in the plugs.  The three-dimensional
structure of the magnetic flux tube can be viewed in \fref{post3d}.
\Fref{fkstm} provides a schematic diagram of the axial magnetic
field of a KSTM; \fref{denff2} provides a conceptual sketch of the
density profile associated with the kinetic stabilizer. At the outer
wall, on the left-side of the figure \fref{denff2}, the magnetic field is at
its lowest value $B=B_{\mathrm{W}}$.  The density of the beam initially
rises in the expander region as we consider regions inward from the
wall. The beam can either be focused on a target in the region around
$B=B_{\mathrm{ks}}$ or unfocused. In the focused case, the density will
peak steeply around $B=B_{\mathrm{ks}}$ and then plummet to a low value
just beyond $B=B_{\mathrm{ks}}$.  The focused kinetic stabilizer ion
beam is designed to reflect around the region of $B_{\mathrm{ks}}$ which
bounds the region of good curvature that exists
between the target and the wall. In the unfocused case the density will
continue to rise until most of the kinetic stabilizer beam has been
reflected by the rising magnetic field. The remnant beam that reaches
this region possess a nearly spatially-isotropic distribution, which
produces a plateau in density. In the plateau region, the density of the
unfocused beam is appreciably higher than for a
focused beam. The electron temperature between the wall and the plateau
region will likely be lower than the ion energy in the injected
kinetic stabilizer. However, the temperature of the electrons
escaping from the central cell will be from $10$-$100$ keV, and thus the
ambipolar potential associated with the effluent flux will be extremely
high. We expect a transition region around a stand-off position,
$B=B_{\mathrm{st}}$, where electron temperature transitions from a
characteristic temperature of the KS region, to the electron temperature
of the core plasma. In this region the plasma density will rise.
%FIG. 2

The density profile and the relation between density the
electric potential will be discussed further in\sref{seckstmmodel} where we
present the details of the model used for the KSTM. In \sref{secmhdstab} we
derive MHD stability relations for the KSTM. In\sref{sectpm} the trapped
particle instability is examined in the context of a KSTM and in\sref{secconc}
we summarize the salient conclusions of this work.

%FIG. 3

\section{KSTM Model \label{seckstmmodel}}

\subsection{Distribution of ions in the kinetic stabilizer beam}

The kinetic stabilizer design consists of an ion beam that is injected
from outside the machine into the expander region of the tandem mirror
where the curvature is positive. The beam of ions is aligned nearly
parallel to the magnetic field. The small pitch-angle of the particles
in the beam corresponds to injection of ions with small
magnetic-moment-per-unit-mass $\mu$ compared to $E_0/B$ (where $E_0$ is the
mean injected-beam-energy-per-unit-mass).  We choose a distribution function
for the ions in the kinetic stabilizer beam such that they possess a
fixed energy equal to $E_0$. The magnetic moment is centered around a value
$\mu_0$ with a narrow range of magnetic moments $\Delta\mu$. The
distribution function, $F$, is chosen to be
\begin{eqnarray}
F(E,\mu) &=& {\Gamma'_{\mathrm{ks}}\over\pi}\delta(E-E_0)
\frac{\Delta\mu}{(\mu-\mu_{\mathrm{T}})^2 + \Delta\mu^2} ~. \label{distribution}
\end{eqnarray}

\noindent Here the quantity $\Gamma'_{\mathrm{ks}}$ is related to the
particle flux per unit magnetic flux $\Gamma_{\mathrm{ks}}$ from a
single end of the tandem mirror as indicated in \eqref{gamma_def}.  
The spread in magnetic moments of the kinetic stabilizer beam is
determined by $\Delta\mu$; we use $\mu_{\mathrm{T}} =
E_0/B_{\mathrm{T}}$ where the subscript $\mathrm{T}$ refers to the
target position; the values of $\Delta\mu$ and $\mu_{\mathrm{T}}$
effects the maximum density at the target. We neglect the ion
electrostatic potential energy under the assumption that the potential
energy is proportional to $T_{\mathrm{eks}}$, the electron temperature
in the kinetic stabilizer region.  The electron temperature is assumed
to be significantly less than the ion beam energy.

 We will treat two
types of kinetic-stabilizer ion-beams.  In the case of a focused beam,
the beam is focused on a target outside of the plug $\mu_{\mathrm{T}}$. 
For the second type of beam, $\mu_{\mathrm{T}}=0$; this will be referred
to as the unfocused case, equivalent to the condition $
\mu_{\mathrm{T}}\ll\Delta\mu$.

The density and pressure in the outer regions of the KSTM are
established by an incoming particle flux at the wall. The subscript
$\mathrm{W}$ refers to the wall position, \it i.e. \rm
$z=z_{\mathrm{W}}$ and $r=r_{\mathrm{W}}$. In terms of the distribution,
the particle flux at the wall \rm $\pi r^2_{\mathrm{W}}  B_{\mathrm{W}}
\Gamma_{\mathrm{ks}}$, is
\begin{eqnarray}
\pi r^2_{\mathrm{W}} B_{\mathrm{W}} \Gamma_{\mathrm{ks}}  &=& \pi r^2_{\mathrm{W}} \int\limits_0^\infty dE\int\limits_0^{E/B_{\mathrm{W}}}d\mu B_{\mathrm{W}} 
F(E,\mu)~,
\\
&=& r_{\mathrm{W}}^2 B_{\mathrm{W}} \Gamma'_{\mathrm{ks}}
\int\limits_0^{E_0/B_{\mathrm{W}}}\ {d\mu\ \Delta\mu\over\left(\mu-\mu_{\mathrm{T}}\right)^2+\Delta\mu^2}.\label{particleflux}
\end{eqnarray}

\noindent We find
\begin{eqnarray}
{\Gamma_{\mathrm{ks}}\over\Gamma'_{\mathrm{ks}}}
\approx {1\over\pi}\int\limits_0^\infty {d\mu\Delta\mu\over\left(\mu-\mu_{\mathrm{T}}\right)^2+\Delta\mu^2} 
= {1\over 2}+{1\over\pi}\,\tan^{-1}\left({\mu_{\mathrm{T}}\over\Delta\mu}
\right)\equiv\chi\left({\mu_{\mathrm{T}}\over\Delta\mu}\right).
\label{gamma_def}
\end{eqnarray}

\noindent For the unfocused case, $\mu_{\mathrm{T}}/\Delta\mu\ll 1$,
$\chi\approx 1/2$, while for the focused case,
$\mu_{\mathrm{T}}/\Delta\mu\gg 1$, $\chi\approx 1$.

\subsection{Density and pressure in the expander due to the kinetic stabilizer beam}

When the effect of the ambipolar potential on the ions is neglected, the density
$n(B)$ is
\begin{equation}
n(B) = \sqrt 2\int\limits_0^\infty d E\int\limits_0^{E/B}
\frac{d\mu B}{\sqrt{E-\mu B}} F(E,\mu).
\end{equation}

\noindent We define the quantity $x=\mu B/E_0$  and use
the distribution function in \eqref{distribution} to express the
density integral as
\begin{eqnarray}
&~&n(B) = \frac{\sqrt{2} \Gamma'_{\mathrm{ks}} B
}{\pi \sqrt{E_0}}
\left( \frac{\Delta \mu B}{E_0}\right)
 \int\limits_0^1 \frac{dx}{\sqrt{1-x}} \frac{1}{(x-B/B_{\mathrm T})^2 
+ (\Delta \mu B/E_0)^2}~,
\\ \nonumber
&~&\xrightarrow[\text{$1\gg\displaystyle{B\over B_{\mathrm{T}}}-1,\ \displaystyle{\Delta\mu B_{\mathrm{T}}\over E_0}\ll 1$}]\ \Gamma'_{\mathrm{ks}}
B\sqrt{\displaystyle{1\over E_0}}\ {\left[\sqrt{\left(1-\displaystyle{B\over B_{\mathrm{T}}}\right)^2+\left(\displaystyle{\Delta\mu B\over E_0}\right)^2}+\left(1-\displaystyle{B\over B_{\mathrm{T}}}\right)\right]^{1/2}\over\sqrt{\left(1-\displaystyle{B\over B_{\mathrm{T}}}\right)^2+\left(\displaystyle{\Delta\mu B\over E_0}\right)^2}}\label{xarrow}~,
\\ \nonumber
&~&\xrightarrow[\text{$1-\displaystyle{B\over B_{\mathrm{T}}}\gg\displaystyle{\Delta\mu B\over E_0}$}]{}
\Gamma'_{\mathrm{ks}}\ \displaystyle\sqrt{2\over E_0}\ \displaystyle{B\over\displaystyle\sqrt{1-\displaystyle{B\over B_{\mathrm{T}}}}}~.
\end{eqnarray}

\noindent The inequality $1\gg B/B_{\mathrm{T}}-1$ is satisfied for all
$B/B_{\mathrm{T}}<1$ when $\Delta\mu B_T/E_0 << 1$. The  last approximation is also accurate
when $|B/B_{\mathrm{T}}-1|\gg\Delta\mu B_{\mathrm{T}}/E_0$, but still
small.

The inequality $\Delta\mu B/E_0\gg 1$, determines a region where the density
asymptotes to a constant value.  We call this region the plateau region, where the density $n_{\mathrm{pla}}$ is
\begin{equation}
n_{\mathrm{pla}} = {2\sqrt 2\,\Gamma'_{\mathrm{ks}} B_{\mathrm{T}}\over\pi\sqrt{E_0}}
\displaystyle{\Delta\mu B_{\mathrm{T}}\over E_0}\doteq {2\over\pi}\,\displaystyle{\Delta\mu B^2_T
\over E_0 B_{\mathrm{W}}} n_{\mathrm{W}}\left(1-\displaystyle{B_{\mathrm{W}}\over B_{\mathrm{T}}}\right)^{1/2}\label{value}~.
\end{equation}
\noindent Here $n_{\mathrm{W}}\equiv n(B_{\mathrm{W}})$ is the density at the wall.

For the focused case the density peaks near the target position
$B=B_{\mathrm{T}}$. At the target the density is given by:
\begin{equation}
n(B_{\mathrm{T}})={\Gamma'_{\mathrm{ks}} B_{\mathrm{T}}\over\sqrt{2E_0}}\left({E_0\over\Delta\mu
B_{\mathrm{T}}}\right)^{1/2} = {n_{\mathrm{W}} B_{\mathrm{T}}\over 2B_{\mathrm{W}}}\left(\displaystyle{E_0\over\Delta\mu
B_{\mathrm{T}}}\right)^{1/2}\left(1-\displaystyle{B_{\mathrm{W}}\over B_{\mathrm{T}}}\right)^{1/2}.\label{peaks}
\end{equation}

\noindent For the unfocused case the density increases monotonically from the wall position to the
plateau region.   We can relate the density at the wall to the density at the entrance to the region of positive curvature, which we will call the kinetic stabilizer position, denoted by the subscript $\mathrm{ks}$. 
Assuming that the axial speed of the beam is small compared to the local thermal spread of its speed, \emph{i.e.} $\Delta\mu B_{\mathrm{ks}}/E_0\ll 1$, then
\begin{equation}
{n\left(B_{\mathrm{ks}}\right)\over n_{\mathrm{W}}} \approx {B_{\mathrm{ks}}\over B_{\mathrm{W}}}
\\
{n_{\mathrm{pla}}\over n_{\mathrm{W}}} \approx {4E_0\over\pi B_{\mathrm{W}}\Delta\mu}\left(1-{B_{\mathrm{W}}\over B_{\mathrm{T}}}\right)^{1/2} ~.
\label{unfocus} 
\end{equation}

\noindent The plateau region arises when $\Delta\mu B\gg E_0$.  By focusing the ion beam to a position $B=B_{\mathrm{T}}$, the
peak density in the kinetic stabilizer region is enhanced compared to
the unfocused case by an approximate factor of $(E_0/\Delta\mu
B_{\mathrm{T}})^{1/2}$. In the plateau region, the density of the
unfocused case is larger than the focused case by a factor of $(E_0/\Delta\mu B_{\mathrm{T}})^{2}$.

In magnetized mirror-confined plasmas the pressure is typically
anisotropic with different values for the pressures $p_\perp$, and
$p_\|$ respectively. The mass of ions in the kinetic stabilizer beam is
taken as $m_{\mathrm{i}} $.   For our choice of distribution function,
the pressures are 
\begin{eqnarray} \label{ptotform}
p(B) &=& p_\|(B)+p_\perp(B) = n(B) m_{\mathrm{i}} E_0 +p_\|(B)/2 ~,
\\
p_\|(B) &=& 2\sqrt 2\,  m_{\mathrm{i}}  \int\limits_0^\infty dE \int\limits_0^{E/B}
d\mu B\sqrt{E-\mu B}~ F(E,\mu) \label{pparform}~,
\\ \label{pperpform}
p_\perp(B)&=&2\, m_{\mathrm{i}}  \int\limits_0^\infty dE\int\limits_0^{E/B}
\frac{d\mu\mu B^2}{\sqrt{2(E-\mu B)}} F(E,\mu)=n(B) m_{\mathrm{i}}  E_0-{p_\|(B)\over 2} ~.
\end{eqnarray}

\noindent The integral for $p_\|(B)$ is
\def\m@th{\mathsurround=0pt} 
\def\n@space{\nulldelimiterspace=0pt \m@th} 
\def\Biggggg#1{{\mbox{$\left#1\vbox to 50pt{}\right.\n@space$}}} 
\begin{eqnarray}
p_\|(B)&&=\frac{2\sqrt 2}{\pi} m_{\mathrm{i}} \Gamma'_{\mathrm{ks}}\, B\sqrt{E_0}\int\limits_0^1
\frac{dx\displaystyle{\Delta\mu B\over E_0}\sqrt{1-x}}{\left(x-\displaystyle{B\over B_{\mathrm{T}}}\right)^2
+\left(\displaystyle{\Delta\mu B\over E_0}\right)^2}~,
\\
&&\mbox{}\xrightarrow[\text{$\displaystyle\Delta\mu = 0$}]\ 2 m_{\mathrm{i}}  \Gamma'_{\mathrm{ks}} \sqrt{2E_0}\left(1-
\frac{B}{B_{\mathrm{T}}}\right)^{1/2}
\theta\left(1-\frac{B}{B_{\mathrm{T}}}\right) ~,
\label{p-parallel}
\end{eqnarray}

\noindent where $\theta(x)$ is the Heaviside step function.  For the focused case where $\Delta\mu B_{\mathrm{T}}/E_0\ll 1$,  the perpendicular pressure
peaks near the target,  $B\approx B_{\mathrm{T}}$, and is approximately
\begin{eqnarray}
p_\perp(B)\cong m_{\mathrm{i}}  E_0 n(B)~,
\end{eqnarray}

\noindent while the parallel pressure remains small. Thus the ratio of pressure at the target to that at the wall is proportional
to the ratio of magnetic fields at these positions
\begin{eqnarray}
{p_\perp(B_{\mathrm{T}})\over p_\perp (B_{\mathrm{W}})}\sim{B_{\mathrm{T}}\over B_{\mathrm{W}}}\left({E_0\over\Delta\mu B_{\mathrm{T}}}\right)^{1/2}.
\end{eqnarray}

\noindent In the plateau region the pressure is isotropic and constant with
\begin{equation}
p_\perp(B) = p_\|(B) = {4\over 3}\sqrt 2\,\pi\,{\Gamma'_{\mathrm{ks}}\over\sqrt{E_0}}
{\Delta\mu B^2_T\over\left[1+\left(\displaystyle{\Delta\mu B_{\mathrm{T}}\over E_0}\right)^2\right]}.
\label{bridge}
\end{equation}

\noindent When $\Delta\mu B_{\mathrm{T}}\gg E_0$, which is the definition of the unfocused case, the pressure in the
plateau region is $p_\perp(B) = p_\|(B) = \sqrt{2}\pi (4/3)
(\Gamma'_{\mathrm{ks}}/\sqrt{E_0}) (E^2_0/\Delta\mu)$.  This pressure is larger than the focused case by a factor of
$ (E_0 / \Delta\mu  B_{\mathrm{T}})^2$.

\section{KSTM MHD Stability Relations \label{secmhdstab}}

During steady-state operation, the power $P$ required to sustain the 
power drain of a single-pass kinetic-stabilizer beam is
\begin{equation}
P = 2\pi m_{\mathrm{i}}  E_0\Gamma_{\mathrm{ks}}\psi_0 ~.
\end{equation}
This expression uses the magnetic flux $\psi_0= \frac{1}{2} B_0 r_0^2$
at the center of the central-cell of the tandem mirror.  The limiting
condition for achieving break-even for reactor engineering is that the
power drain from the kinetic stabilizer beams should not exceed  the
alpha particle energy production rate of the central-cell by more than
some factor $\eta$:
\begin{equation}
\eta \pi r_0^2 m_{\mathrm{i}}  E_0 B_0\Gamma_{\mathrm{ks}} <
\frac{3\pi r_0^2 L_{\mathrm{c}} T_{\mathrm{c}} n_{\mathrm{c}} ^2}
{2\left<n\tau\right>_{\mathrm{fus}}}\label{powerbalance} ~.
\end{equation}

\noindent where $\eta=1$ is the natural choice.  The temperature in this
expression is a sum of electron and ion temperatures $ T_{\mathrm{c}} =
T_{\mathrm{ec}} + T_{\mathrm{ic}}$ and the subscript c refers to the
central-cell.  The Lawson criterion
$\langle{n\tau}\rangle_{\mathrm{fus}} = 2\times 10^{20}\,$m$^{-3}$
should hold for our model KSTM.   For MHD stability, we require
\begin{equation}
\int\limits_{\mathrm{ks}} dz\left(p_\|+p_\perp\right) r^3 \frac{d^2 r}{dz^2}>
\int\limits_{\mathrm{plug~+~central~cell}} dz\left(p_\|+p_\perp\right) r^3\frac{d^2 r}{dz^2}.
\label{ksplugbalance}
\end{equation}

\noindent To evaluate the MHD stability criterion in 
\eqref{ksplugbalance} we constrain the field line radius $r(z)$ to have
as large as possible variation within the kinetic stabilizer region that
is compatible with the paraxial approximation. Thus, we take field-line
radius $r(z)$ must satisfy
\begin{equation}
r\,{d^2 r(z)\over dz^2} = r(z){d\over dr}\left({dr(z)\over dz}\right)^2 = {\sigma_{\mathrm{p}}\over 2}.\label{mp_axial}
\end{equation}
We will refer to \eqref{mp_axial} as the marginal paraxial
constraint. For optimal stability, the constant
$\sigma_{\mathrm{p}}$ in the the marginal paraxial constraint should be
chosen to maximize the integrand on the left hand side of
\eqref{ksplugbalance}, subject to the validity of the paraxial
approximation for the field-line curvature; in this work we use
$\sigma_{\mathrm{p}} = 1$ while acknowledging that additional study is
needed to determine an optimal value.

Solving \eqref{mp_axial} for $r(z)$ yields
\begin{eqnarray}\label{zzrmks}
z-z_{\mathrm{ks}}&=& \int\limits_{r_{\mathrm{ks}}}^r  
\frac{dr}{\left[\sigma_{\mathrm{p}}\,\ln\left(\displaystyle\frac{r}{r_{\mathrm{ks}}}\right) + 
\left(\displaystyle\frac{dr(z_{\mathrm{ks}})}{dz}\right)^2\right]^{1/2}}
\\
&\le&
\displaystyle{r_{\mathrm{W}}\over\sqrt{\sigma_{\mathrm{p}}}}
\int\limits^{r/r_{\mathrm{W}}}_{r_{\mathrm{ks}}/r_{\mathrm{W}}} \displaystyle{dy\over\left[\ln\left(\displaystyle{r_{\mathrm{W}}\over r_{\mathrm{ks}}}y\right)\right]^{1/2}}. \nonumber
\end{eqnarray} 

\noindent When $dr(z_{\mathrm{ks}})/dz=0$, the inequality in 
\eqref{zzrmks} is an equality.  Since the choice of
$dr(z_{\mathrm{ks}})/dz=0$ maximizes the favorable MHD response, this
field-line slope is used in our model for KSTM field-lines.

\subsection{Stability integral in the plug}

We model the field-line radius in the plug region with the form
\begin{eqnarray}
r(z) = r_{\mathrm{mxp}}\Bigg({1+x_{\mathrm{pl}}\over 2} + {1-x_{\mathrm{pl}}\over 2}
\cos\left(\frac{2\pi(z-L_{\mathrm{c}})}{ L_{\mathrm{pl}} }\right)\Bigg)~,
\label{seventeen}
\end{eqnarray}
where $x_{\mathrm{pl}} =
r_{\mathrm{mnp}}/r_{\mathrm{mxp}}$. The radii  $r_{\mathrm{mxp}}$ and
$r_{\mathrm{mnp}}$ are the maximum and
minimum radii of the plug region, and
$0<|z-L_{\mathrm{c}}|<L_{\mathrm{pl}}$. We define the composite pressure 
$p = (p_\perp + p_\|)/2$.   We also define
\begin{eqnarray}
\overline p_{\mathrm{pl}}  = {1\over 2L_{\mathrm{pl}}} \int\limits_{L_{\mathrm{c}}}^{L_{\mathrm{c}}+L_{\mathrm{pl}}} dz\left(p_\perp + p_\|\right) = {1\over 2L_{\mathrm{pl}}} \int\limits_{{\mathrm{pl}}} dz ~p ~.
\end{eqnarray}

\noindent The dimensionless plasma parameter in this region is defined $\overline\beta_{\mathrm{pl}}= 2\mu_0\overline p_{\mathrm{pl}}
/B^2_{\mathrm{mnp}}$, where $B_{\mathrm{mnp}}$ is the magnetic field at the minimum field-line radius in the plug region.  We define $I_{\mathrm{pl}}$ to be the stability integral over
the plug region for the particular magnetic field assumed in \eqref{seventeen}. Thus we need
to evaluate
\begin{eqnarray}
I_{\mathrm{pl}} =  -2\int\limits_{\mathrm{pl}} dz\overline p_{\mathrm{pl}}  r^3\frac{d^2 r}{dz^2} ~.
\label{mflr}
\end{eqnarray}
\noindent We indicate an approximation of $I_{\mathrm{pl}}$ with a prime. We evaluate $I'_{\mathrm{pl}}$, as the
approximate stability integral $I_{\mathrm{pl}}$ when the pressure is taken as
isotropic in the plug with a pressure $\overline p_{\mathrm{pl}}$. We obtain
\begin{eqnarray} \label{ipl}
I'_{\mathrm{pl}} &=&\frac{3\pi^2\overline p_{\mathrm{pl}} }{4L_{\mathrm{pl}}} r_{\mathrm{mxp}}^4
\left(1-x_{\mathrm{pl}}^2\right)^2\Bigg(1+\frac{1}{4}\left(\frac{1-x_{\mathrm{pl}}}
{1+x_{\mathrm{pl}}}\right)^2\Bigg).
\end{eqnarray}

\noindent  In this expression we have used the paraxial approximation for the magnetic flux
$r^2_{\mathrm{mxp}} B_{\mathrm{mnp}}\approx B_0 r^2_0$.  For simplicity we have neglected
the small factor
$(1-x_{\mathrm{pl}})^2/(2\left(1+x_{\mathrm{pl}}\right))^2$ in the
estimate $I'_{\mathrm{pl}}$.  We define $\alpha_{\mathrm{pl}} $ as a
dimensionless, order-unity parameter that is the ratio of the exact to
the estimated value of the stability integral
\begin{eqnarray} 
\alpha_{\mathrm{pl}}  = I_{\mathrm{pl}}/I'_{\mathrm{pl}} .
\end{eqnarray}
\noindent The factor $\alpha_{\mathrm{pl}}$ may be less than unity (\it e.g.~\rm due to the inclusion
of compressibility effects in the plug cell and improved design of the plug, \it etc.\rm) and
such improvement gives some flexibility in estimating a window of operation for the
KSTM.

The contribution to MHD stability from the central-cell must also be considered
and is treated in the same way as the plug. We use a shape of the same form as as the plug region described in
\eqref{seventeen}, and central-cell parameters $r_{\mathrm{mxp}} \to r_0$,
$L_{\mathrm{pl}}\to 2 L_{\mathrm{c}}$, $z-L_{\mathrm{c}}\to z$, $x_{\mathrm{pl}}\to x_{\mathrm{c}} = r_0/r_{\mathrm{mn}}$.   Thus
\begin{eqnarray}
r(z) = r_{0}\Bigg({1+x_{\mathrm{c}}\over 2} + {1-x_{\mathrm{c}}\over 2}
\cos\left(\frac{2\pi z}{ L_{\mathrm{c}} }\right)\Bigg) ~.
\label{seventeenagain}
\end{eqnarray}

\noindent The central-cell stability contribution will be used below.

\subsection{Stability integral in the expander region}

For our KSTM model there is no contribution to the stability integral from
the intermediate region between the outlet of the plug (defined by the
maximum of the magnetic field) and the beginning of the positive-curvature kinetic stabilizer
region at $z=z_{\mathrm{ks}}$. The stability contribution of this
intermediate expander region would vanish if either $d^2 r/dz^2$ or the
kinetic pressure were negligible.

The MHD contribution of the stability drive integral from the
expander region is
\begin{equation}
I_{\mathrm{ks}} = \int\limits_{z_{\mathrm{ks}}}^{z_{\mathrm{W}}} dz\left(p_\|+p_\perp\right)r^3~\kappa~.
\label{mhd}
\end{equation} 

\noindent In this expression $\kappa$ is the field-line curvature, 
\begin{equation}
\kappa = {\displaystyle{d^2r\over dz^2}\over\left[1+\left(\displaystyle{dr\over dz}\right)^2\right]^{3/2}} \approx \frac{d^2r}{dz^2}~.
\label{express}
\end{equation}
\noindent The paraxial approximation of the curvature is given in the
right-most term of \eqref{express}.

To establish a base-case for scaling our results, we consider the
unfocused kinetic stabilizer beam ($\Delta\mu=0$), and use the
paraxial approximation for the curvature. Using the fact that magnetic flux
is conserved, $Br^2=B_0 r_0^2$, the expressions for the
pressures from \eqref{pparform} and  \eqref{pperpform}, and the
solution $r(z)$ from \eqref{zzrmks}, we find
\begin{eqnarray}\label{b}
I_{\mathrm{ks}} &=& \int\limits_{z_{\mathrm{ks}}}^{z_{\mathrm{W}}} dz\left(p_\|+p_\perp\right)\bigg|_{\Delta\mu=0}\ r^3 
{d^2r\over dz^2} 
\\ \nonumber
&=& 2m_{\mathrm{i}} \sqrt{2\sigma_{\mathrm{p}} E_0}\ B_0 r_0^2\Gamma'_{\mathrm{ks}} 
r_{\mathrm{W}}\int\limits^1_{r_{\mathrm{ks}}/r_{\mathrm{W}}}
{dy\over\sqrt{\ln\left( \displaystyle{r_{\mathrm{W}}\over r_{\mathrm{ks}}}y \right)}} ~,
\end{eqnarray}

\noindent where we have used the slope $dr_{\mathrm{ks}}/dz = 0$.  We approximate the stability integral in the expander region
as
\begin{eqnarray}
I'_{\mathrm{ks}} &=& 2m_{\mathrm{i}}\sqrt{\sigma_{\mathrm{p}}2E_0}\ B_0 r_0^2
\Gamma'_{\mathrm{ks}}r_{\mathrm{W}} ~.
\end{eqnarray}

\noindent We define the ratio of the exact to the approximate stability integrals 
\begin{eqnarray}
\alpha_{\mathrm{ks}} &=& {I_{\mathrm{ks}}\over I'_{\mathrm{ks}}}. \label{c}
\end{eqnarray}
\noindent In \eqref{b} the use of the paraxial approximation for
the curvature allows $I_{\mathrm{ks}}$ to be made arbitrarily large by
increasing $\sigma_{\mathrm{p}}$.  However, as $\sigma_{\mathrm{p}}$
increases, the paraxial approximation is eventually violated. To verify consistency between
$I_{\mathrm{ks}}$  and the approximate
paraxial approximation, we choose $\sigma_{\mathrm{p}}=1$ and
compare results for the non-paraxial expression for the field-line curvature with the results of the
paraxial approximation. We use the paraxial solution for z from 
\eqref{zzrmks} and the more accurate form of the curvature $\kappa$ to
calculate the integral
\begin{eqnarray}
{\sigma_{\mathrm{p}}\over r_{\mathrm{W}}}\int\limits^{z_{\mathrm{W}}}_{z_{\mathrm{ks}}}
{dzr \kappa \over\left(1+
\left(\displaystyle{dr\over dz}\right)^2\right)^{3/2}}
=\int\limits^1_{r_{\mathrm{ks}}/r_{\mathrm{W}}}
\displaystyle{dy\over\sqrt{\ln\left(\displaystyle{r_{\mathrm{W}}\over r_{\mathrm{ks}}}y\right)}}
\frac{1}{\Bigg(1+\sigma_{\mathrm{p}}\ln\left(\displaystyle{r_{\mathrm{W}}\over r_{\mathrm{ks}}}y\right)\Bigg)^{3/2}}~.~~~~~~
\end{eqnarray} 

\noindent For the unfocused case \eqref{b} becomes
\begin{eqnarray}
I_{\mathrm{ks}\,0}\equiv\int\limits_{z_{\mathrm{ks}}}^{z_{\mathrm{W}}}
dz r^2\kappa\left(p_\|+p_\perp\right)
\bigg|_{\Delta\mu=0} = I'_{\mathrm{ks}}\int\limits^1_{r_{\mathrm{ks}}/r_{\mathrm{W}}}
{ dy\over\sqrt{\ln\left(\displaystyle{r_{\mathrm{W}}\over r_{\mathrm{ks}}}y\right)}
\Bigg(1+\sigma_{\mathrm{p}}\ln\left(\displaystyle{r_{\mathrm{W}}\over r_{\mathrm{ks}}}y\right)\Bigg)^{3/2}}~.\label{e}~~~~~~
\end{eqnarray}
%FIG. 4
\begin{figure}[H]
\includegraphics[scale=.9]{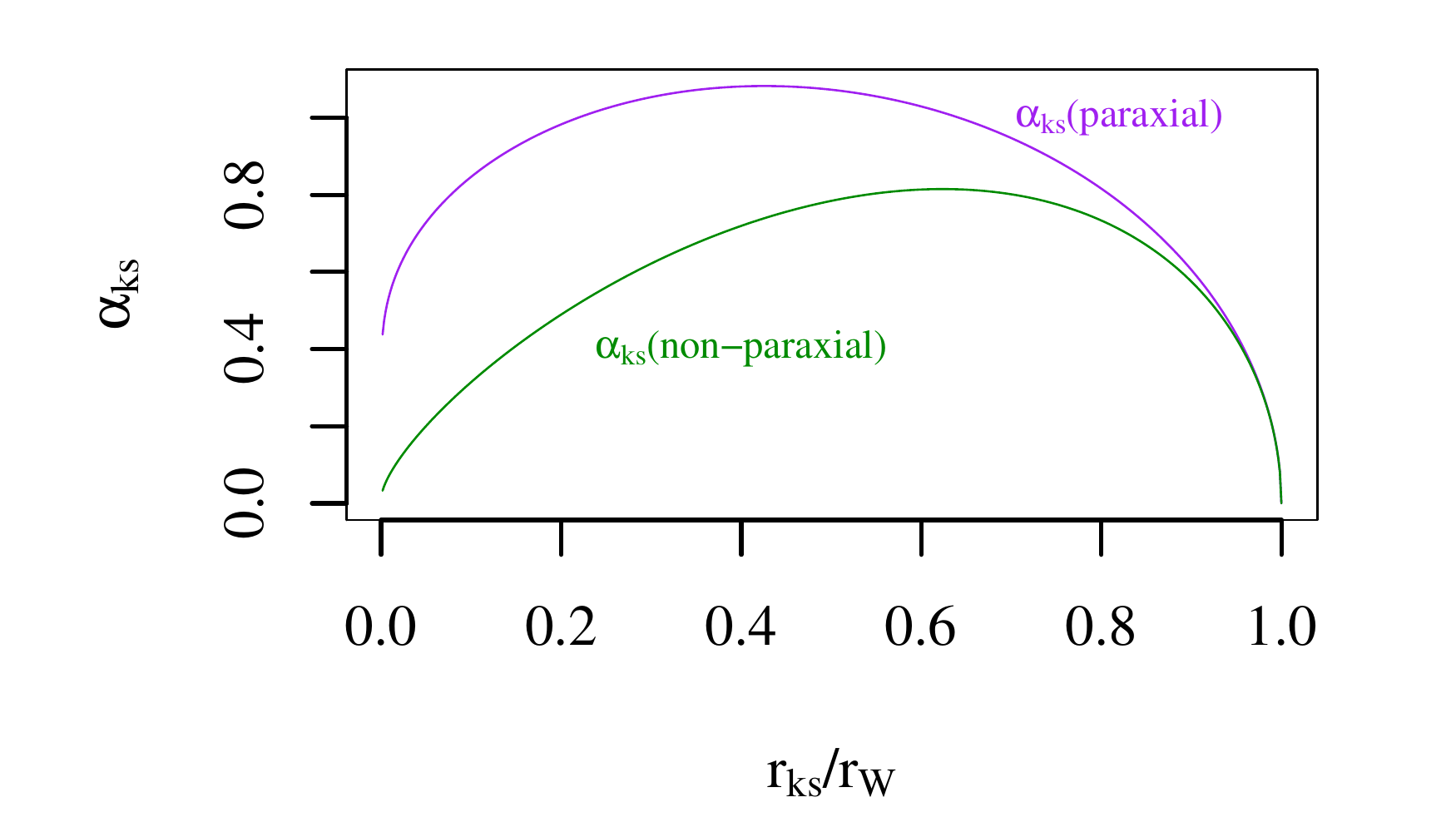}
\caption{The quantity $\alpha_{\mathrm{ks}} =
I_{\mathrm{ks}}/I'_{\mathrm{ks}}$ evaluated for an unfocused kinetic
stabilizer beam from \eqref{e} using $\sigma_{\mathrm{p}}=1$.}
\label{nofo3}
\end{figure}

\Fref{nofo3} shows the comparative accuracy of the paraxial approximation
and the more accurate non-paraxial curvature for
$\alpha_{\mathrm{ks}} = I_{\mathrm{ks}}/I'_{\mathrm{ks}}$ as a function
of $x = r_{\mathrm{ks}}/r_{\mathrm{W}}$. The maxima are
$\alpha_{\mathrm{ks}}=1.08$ for $x_{\mathrm{ks}}=0.43$ for the paraxial
form and $\alpha_{\mathrm{ks}}=0.815$ for $x_{\mathrm{ks}}=0.63$ for the
non-paraxial form. There is some quantitative discrepancy, which indicates that
$\sigma_{\mathrm{p}}=1$ may be the largest value that can be used in the expression for $I_{\mathrm{ks}}$
  to give reliable results.

Larger values of $I_{\mathrm{ks}}$ can be obtained by focusing the beam
so that the target is at the kinetic stabilizer point $B_{\mathrm{T}}=B_{\mathrm{ks}}$. When the beam is focused
on this location, a logarithmic divergence arises  in $I_{\mathrm{ks}}$,
producing arbitrarily strong MHD stabilization. For a beam with thermal
spread (finite $\Delta\mu$),  the magnitude of the
stability integral $I_{\mathrm{ks}}$ is bounded. To evaluate the
stability integral in this case we use an approximate form for the
pressure:
\begin{eqnarray} \nonumber
{2p\over B m_{\mathrm{i}} \sqrt{E_0}\,\Gamma'_{\mathrm{ks}}} = 
\\
{\left[\sqrt{\left(1-\displaystyle{B\over
B_{\mathrm{T}}}\right)^2+\left(\displaystyle{\Delta\mu B\over E_0}\right)^2}+1-\displaystyle{B\over
B_{\mathrm{T}}}\right]^{1/2}\over\sqrt{\left(1-\displaystyle{B\over B_{\mathrm{T}}}\right)^2
+\left(\displaystyle{\Delta\mu B\over E_0}\right)^2}}+\sqrt{1-\displaystyle{B\over B_{\mathrm{T}}}}
\theta(1-\displaystyle{B\over B_{\mathrm{T}}} )
~.
\label{pressure}
\end{eqnarray}

\noindent This is valid if $1\gg B/B_{\mathrm{T}}-1$, a condition that 
is always satisfied when the right-hand side is negative and $\Delta \mu B_T/E_0 \ll1$.
Again using the condition that magnetic flux is conserved, so $B/B_{\mathrm{ks}}=(r_{\mathrm{ks}}/r)^2$, and setting the
target position T at the kinetic stabilizer position KS, we obtain a
stability integral $I_{\mathrm{fo}}\equiv I_{\mathrm{ks}}$ for the
focused case
\begin{eqnarray}
I_{\mathrm{fo}} = I'_{\mathrm{ks}}~g \left(\frac{r_{\mathrm{ks}} }{r_{\mathrm{W}}},\frac{\Delta \mu B_{\mathrm{ks}}}{E_0} \right)~,
\end{eqnarray}

\begin{eqnarray}
g(y,\epsilon)=\int\limits^1_{y} \frac{dx}{\sqrt{\ln\left(\displaystyle{x\over y}\right)}\Bigg(1+\sigma_{\mathrm{p}}\ln
\left(\displaystyle{x\over y}\right)\Bigg)^{3/2}} 
\\  \nonumber
[\theta\left(1-\left(\displaystyle{\frac{y}{x}}\right)^2\right) \sqrt{2\left(1-\left(\displaystyle{\frac{y}{x}}\right)^2\right)}
+
\\  \nonumber
{\Bigg(\sqrt{\left(1-\left(\displaystyle{y\over x}\right)^2\Bigg)^2+
\Bigg(\epsilon \Bigg(\displaystyle{y\over x}\right)^2\Bigg)^2}+\sqrt{1-\left(\displaystyle{y\over x}\right)^2}\Bigg)^{1/2}
\over\sqrt{\Bigg(1-\left(\displaystyle{y\over x}\right)^2\Bigg)^2
+\Bigg(\epsilon \left(\displaystyle{y\over x }\right)^2\Bigg)^2}} ]
%\\ \nonumber
%&~&+\theta\left(1-\left(\displaystyle{\frac{y}{x}}\right)^2\right) \sqrt{2\left(1-\left(\displaystyle{\frac{y}{x}}\right)^2\right)}
%\right)~.\nonumber
\end{eqnarray}

\begin{figure}[H]
\includegraphics[scale=.9]{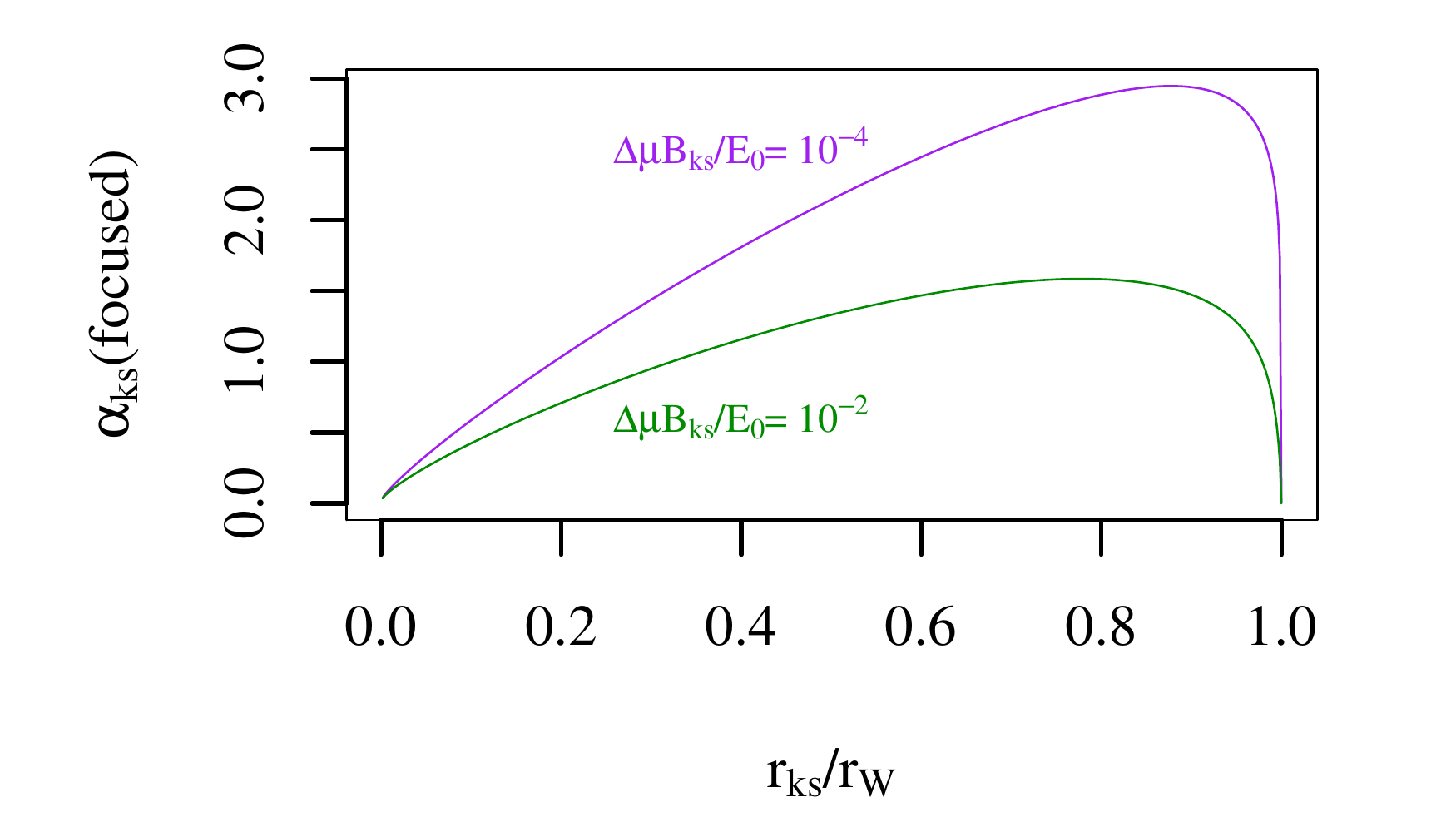}
\caption{The quantity $\alpha_{\mathrm{ks}} = I_{\mathrm{ks}}/I'_{\mathrm{ks}}$ in the focused case, evaluated at two different beam widths [\eqref{b}].}
\label{foyay4}
\end{figure}

\noindent A plot of $\alpha_{\mathrm{ks}} = I'_{\mathrm{fo}}/I'_{\mathrm{ks}}$ as
a function of $r_{\mathrm{ks}}/r_{\mathrm{W}}$ is shown in 
\fref{foyay4} for the focused case where ($\Delta\mu B_{\mathrm{ks}}/E_0= 10^{-2},
10^{-4}$). As
$\Delta\mu B_{\mathrm{ks}}/E_0$ decreases, the optimal MHD response is
found closer to the wall; at small $\Delta\mu B_{\mathrm{ks}}/E_0$, the
value of this optimal response is given by $0.2+0.7
\log_{10}(E_0/\Delta\mu B_{\mathrm{ks}})$. For $\Delta\mu
B_{\mathrm{ks}}/E_0=10^{-4}$ this leads to $\alpha_{\mathrm{ks}}\approx
3$.
%FIG. 5

\subsection{Operating regime scaling laws}

For our machine to be stable we require $M_{\mathrm{ks}}\equiv
(I_{\mathrm{pl}}+I_{\mathrm{c}})/I_{\mathrm{ks}}<1$ or
\begin{eqnarray}\label{machine}
1>M_{\mathrm{ks}} &=& {3\pi^2 r^2_0\left[\alpha_{\mathrm{pl}} \overline p_{\mathrm{pl}} 
\displaystyle{\left(1-x^2_{\mathrm{pl}}\right)^2\over L_{\mathrm{pl}}}\left(\displaystyle{r^2_{\mathrm{mxp}} 
\over r^2_0}\right)^2+\alpha_{\mathrm{c}} n_{\mathrm{c}}  T_{\mathrm{c}}\displaystyle{\left(1-x^2_c\right)^2\over 4L_{\mathrm{c}}}\right]\over
8\alpha_{\mathrm{ks}}\sqrt{\sigma_{\mathrm{p}} 2E_0}\ r_{\mathrm{W}} m_{\mathrm{i}}  B_0\Gamma'_{\mathrm{ks}}}
\\ \nonumber
&=&\ {3\pi^2 r^2_0\ B_0^2\over 16\alpha_{\mathrm{ks}}\ r_{\mathrm{W}}\sqrt{2E_0\sigma_{\mathrm{p}}}\, \mu_0 m_{\mathrm{i}}  B_0\Gamma'_{\mathrm{ks}}}
\left[{\alpha_{\mathrm{pl}}\,\overline\beta_{\mathrm{pl}}\over L_{\mathrm{pl}}}\left(1-x^2_{\mathrm{pl}}\right)^2
+\alpha_{\mathrm{c}}
\displaystyle{\overline\beta_c\left(1-x_{\mathrm{c}}^2\right)^2\over 4 L_{\mathrm{c}}}\right].
\end{eqnarray} 
\noindent where $x_{\mathrm{c}}=r_0/r_{\mathrm{mxp}}$,
$T_{\mathrm{c}}=T_{\mathrm{ec}}+T_{\mathrm{ic}}$.  We shall call
$M_{\mathrm{ks}}$ the MHD marginality parameter.  The stability
condition of \eqref{machine} uses the approximations for the
stability drive integrals in the plug given by ~\eqref{ipl}  and in
the kinetic stabilizer region given by \eqref{b}.  The MHD drive
from the central-cell uses the same functional forms for the magnetic
field's axial variation as the plug region, and employs the appropriate
changes in lengths and mirror ratio. The total length of the
central-cell region is $2L_{\mathrm{c}}$. In this stability criterion
the central-cell drive is integrated over only half of the central-cell.
The quantity $\alpha_{\mathrm{c}}$ is the ratio of the actual MHD drive
to the scaled MHD drive in the central-cell.

From the power-balance relation \eqref{powerbalance} we find
\begin{eqnarray}
{\eta 3\pi r_0^2 L_{\mathrm{c}} T_{\mathrm{c}} n_{\mathrm{c}} ^2\over 2\left<n\tau\right>_{\mathrm{fus}}\chi\left(\displaystyle{\mu_{\mathrm{T}}\over\Delta\mu}\right)}
&>&\pi r_0^2 E_0 B_0 m_{\mathrm{i}} \Gamma'_{\mathrm{ks}}\nonumber\\
&>&{3\pi^3\sqrt{2 E_0}\, B^2_0 r^4_0\over 32\,\sqrt{\sigma_{\mathrm{p}}}\,\alpha_{\mathrm{ks}} 
r_{\mathrm{W}}\mu_0}\left({\alpha_{\mathrm{pl}} \overline\beta_{\mathrm{pl}}\over L_{\mathrm{pl}}}\left(1- x_{\mathrm{pl}}^2\right)^2 +
{\alpha_{\mathrm{c}}\overline\beta_{\mathrm{c}} \left(1-x_{\mathrm{c}}^2\right)^2\over 4L_{\mathrm{c}}}\right).\label{powerbalance2}~~~~~~
\end{eqnarray}

\noindent Thus the condition that the system be MHD stable with an acceptable energy
loss from the kinetic stabilizer beam is
\begin{eqnarray}
{r_{\mathrm{W}}\over r_0} &=& \sqrt{\displaystyle{B_0\over B_{\mathrm{W}}}}
\\ \nonumber
&>&\displaystyle{\pi^2 r_0\over 80 \sqrt{\sigma_{\mathrm{p}}}
n_{\mathrm{c}} L_{\mathrm{c}} L_{\mathrm{pl}} }
{\left<n\tau\right>_{\mathrm{fus}}\sqrt{2E_0} ~ \chi\left({\mu_{\mathrm{T}}\over\Delta\mu}\right)
\over\alpha_{\mathrm{ks}} \eta} \nonumber
 \\ \nonumber
&~&\left(10 \alpha_{\mathrm{pl}} \left(1-x^2_{\mathrm{pl}}\right)^2{\overline\beta_{\mathrm{pl}}\over\overline\beta_{\mathrm{c}}}
+\frac{10}{4}\alpha_{\mathrm{c}}\displaystyle{L_{\mathrm{pl}}\over L_{\mathrm{c}}}\left(1-x^2_{\mathrm{c}} \right)^2\right)
\label{pp}~.
\end{eqnarray}

\noindent The plasma beta is $\beta_{\mathrm{c}}= n T_{\mathrm{c}}  2\mu_0 /B_0^2$.  We use
\begin{eqnarray}
n_{\mathrm{c}}^{-1}(\mathrm{m}^{-3}) &=& 
4.5 \cdot 10^{-21}
~ \frac{T_{\mathrm{c}}(100~\mathrm{keV})}{\beta_{\mathrm{c}}} 
\left(\frac{3}{B_0(\mathrm{T})}\right)^2 ~.
\end{eqnarray}

\noindent Normalizing \eqref{pp} to a reference
parameter of approximately $100\,$MW fusion power production, we find
\begin{eqnarray}
{r_{\mathrm{W}}\over r_0} = \sqrt{\displaystyle{B_0\over B_{\mathrm{W}}}}
>70.{\lambda_{\mathrm{c}}\sqrt{E_0
(\displaystyle{\mathrm{keV}})}\over r_0(\mathrm{m})}\left(\displaystyle{T_{\mathrm{c}}(100\,{\mathrm{keV}})
\over\beta_c}\right)\left(\displaystyle{3\over B_0(\mathrm{T})}\right)^2\label{fpower}~.
\end{eqnarray}

\noindent Here $\lambda_{\mathrm{c}}$ is a dimensionless parameter that characterizes the machine
\begin{eqnarray} \label{lamcdef}
\lambda_{\mathrm{c}}&\equiv&{\chi\left({\mu_{\mathrm{T}}\over\Delta\mu}\right)\over{\alpha_{\mathrm{ks}} \eta}
\sqrt{\sigma_{\mathrm{p}}}}\left(\displaystyle{100 r_0\over
L_{\mathrm{c}}}\right)\left(\displaystyle{5r_0\over L_{\mathrm{pl}}}\right)\left(\displaystyle{2\over m_{\mathrm{i}} (a)}\right)^{1/2}
\\
&~&\left(10 \alpha_{\mathrm{pl}} \left(1-x^2_{\mathrm{pl}}\right)^2{\overline\beta_{\mathrm{pl}}\over\overline\beta_{\mathrm{c}}}
+2.50\alpha_{\mathrm{c}}\displaystyle{L_{\mathrm{pl}}\over L_{\mathrm{c}}}\left(1-x^2_{\mathrm{c}} \right)^2\right) .\nonumber
\end{eqnarray}

\noindent The injected ion mass $m_{\mathrm{i}} (a)$ in the kinetic
stabilizer beam is in atomic units so that $m_{\mathrm{i}} =2$ for
deuterium.  The Lawson number $\langle n\tau \rangle_{\mathrm{fus}}
=2\times10^{20}\,$m$^{-3}$ is used to characterize the alpha particle
power production; the terms in each of the parentheses is estimated to
be order unity and thus $\lambda_{\mathrm{c}}\simeq 1$ is a
characteristic estimate.

The scaling law \eqref{fpower} indicates whether a system with a given set of
parameters can be MHD stable with an acceptable energy loss from the
kinetic stabilizer beam.  We choose a reasonable set of parameters to
test the MHD stability:
\begin{eqnarray}
B_0(\mathrm{T}) &=& 3 \mathrm{T},\ T_{\mathrm{c}}(100) = 100\,\mathrm{keV},\ r_0 = {1\over\sqrt 3}\ \mathrm{m},
\nonumber\\ 
L_{\mathrm{c}} &=& 
{100\over\sqrt 3}\ \mathrm{m}, L_{\mathrm{pl}} = 5r_0 = {5\over\sqrt 3}\, \mathrm{m},
\nonumber\\
\overline\beta_{\mathrm{pl}} &=& {\beta_c\over 5} = \alpha_{\mathrm{ks}} = \alpha_{\mathrm{c}} 
= 2 \alpha_{\mathrm{pl}}  = 1,\
\nonumber\\
 x^2_{\mathrm{pl}} &=& 0.5,\ x_{\mathrm{c}}^2 = 0.1,\ m_{\mathrm{i}} (a)=2,\ \eta=1.
\label{thirtyone}~~~~~~~~
\end{eqnarray}

\noindent With this choice of parameters we find
$\lambda_{\mathrm{c}}=0.175$. We find it necessary to have
$B_{\mathrm{W}} < 66$ gauss when $E_0=1$ keV or  $B_{\mathrm{W}} < 660$
gauss when $E_0=100$ eV. For the parameters in \eqref{thirtyone}
the instability drive of the central-cell is 0.4 times the drive from
the plug. If the drive from the plug could be reduced or eliminated, the
drive in the central-cell alone would require
$B_{\mathrm{W}}<7.9\beta_{\mathrm{c}}^2 10^2\,$ gauss for $E_0=1$ keV (or
$0.79\beta_{\mathrm{c}}^2 $ T if $E_0=100$ eV). Hence significant
improvement could be achieved if the drive from the plug could be
neutralized. For example, the plug drive could be neutralized if the
particle species in the plug were sufficiently energetic \cite{krall66}.

In summary we find that an expander whose magnetic field at the wall
could be as large as the order of several 100 gauss would enable MHD stability
to be achieved by kinetic stabilizer beams of several hundred eV, in
combination with 3~T central-cell magnetic field and
$\alpha_{\mathrm{pl}}\approx 0.5$. The detailed scaling is indicated in
\eqref{fpower} and \eqref{lamcdef}.

\subsection{Beta Limitation and Adiabaticity Limits \label{secbetalim}}

An additional facet of the kinetic stabilizer ion plasma beta should be
considered.  When $\beta$ becomes larger than unity, typically the
field-lines cannot collimate the plasma because self-consistent effects
drive the plasma across field-lines \cite{bbaa}. For the injected
kinetic stabilizer ions, the local $\beta_{\mathrm{ks}}=2 \mu_0 p/B^2$
must be less than unity along the imposed field-lines.  From 
\eqref{pressure} we find that in the kinetic stabilizer region
\begin{eqnarray}
\beta_{\mathrm{ks}} &=& \frac{2 \mu_0 p}{B^2} \nonumber
\\ &=& \frac{\mu_0}{B} m_i \sqrt{E_0} \Gamma'_{\mathrm{ks}}
\left[\frac{\left[ \left[(1-b_{\mathrm{T}})^2 + \epsilon_{\mathrm{ks}}^2 b_{\mathrm{ks}}^2\right]^{1/2} 
+1-b_{\mathrm{T}}\right]^{1/2}}{\left[(1-b_{\mathrm{T}})^2 + \epsilon_{\mathrm{ks}}^2 b_{\mathrm{ks}}^2\right]^{1/2}}
 + 2(1-b_{\mathrm{T}})^{1/2} \right] ~,
\label{newbetaks}~~~~~
\end{eqnarray}

\noindent Expressing $\Gamma'_{\mathrm{ks}}$ in terms of
$\beta_{\mathrm{ks}}$, we find that the MHD stability condition from
\eqref{machine} and \eqref{newbetaks} can be expressed as
\begin{eqnarray} \nonumber
&~&1> M_{\mathrm{ks}} \equiv \frac{3 \pi^2}{64 \sqrt{2}} \frac{r_0^2}{r_{\mathrm{W}}^2} 
\frac{B_0}{B\sqrt{\sigma_{\mathrm{p}}} } 
\frac{r_{\mathrm{W}}}{L_{\mathrm{c}}} 
\frac{\alpha_{\mathrm{c}} \overline\beta_{\mathrm{c}}}{\beta_{\mathrm{ks}}} (1-x_{\mathrm{c}}^2)^2
\left[ 1+ \frac{4\alpha_{\mathrm{pl}}}{\alpha_{\mathrm{c}}} \frac{L_{\mathrm{c}}}{L_{\mathrm{pl}} }
\frac{\beta_{\mathrm{pl}} }{\beta_{\mathrm{c}} } \frac{(1-x_{\mathrm{pl}}^2)^2 }{(1-x_{\mathrm{c}}^2)^2}\right]
\\
&~& 
\left[
\frac{\left[ \left[(1-b_{\mathrm{T}})^2 + \epsilon_{\mathrm{ks}}^2 b_{\mathrm{ks}}^2\right]^{1/2} 
+1-b_{\mathrm{T}}\right]^{1/2} }
{\left[(1-b_{\mathrm{T}})^2 + \epsilon_{\mathrm{ks}}^2 b_{\mathrm{ks}}^2\right]^{1/2}}
 + \sqrt{2 (1-b_{\mathrm{T}})} \right] .~~~
 \label{mksineq}~~~~
\end{eqnarray}

\noindent Here we again use the conservation of flux $(r_0^2 /
r_{\mathrm{W}}^2)(B_0/B) = (B_{\mathrm{W}} /B)$. If the inequality has
to be satisfied at every position, \eqref{mksineq} becomes
\begin{eqnarray}
1&>& M_{\mathrm{ks}} > \frac{0.33 \alpha_{\mathrm{c}}}{\sqrt{\sigma_{\mathrm{p}}} }
(1-x_{\mathrm{c}}^2)^2 \beta_{\mathrm{c}}
g M_{\mathrm{X}}~,
\\
g&=& \left[ 1+ \frac{4\alpha_{\mathrm{pl}}}{\alpha_{\mathrm{c}}} \frac{L_{\mathrm{c}}}{L_{\mathrm{pl}} }
 \frac{\beta_{\mathrm{pl}} }{\beta_{\mathrm{c}} } \frac{(1-x_{\mathrm{pl}}^2)^2 }{(1-x_{\mathrm{c}}^2)^2}\right] ~.
\end{eqnarray}

\noindent Here the expression $M_{\mathrm{X}}$ is the maximum with
respect to $B$ of the function
\begin{eqnarray}
 M_{\mathrm{X}}= \mathrm{Max} \left[  \frac{B_{\mathrm{W}} }{B} \left[ 
 \frac{\left[ \left[(1-b_{\mathrm{T}})^2 + \epsilon_{\mathrm{ks}}^2 b_{\mathrm{ks}}^2\right]^{1/2} 
+1-b_{\mathrm{T}}\right]^{1/2}}{\left[(1-b_{\mathrm{T}})^2 + \epsilon_{\mathrm{ks}}^2 b_{\mathrm{ks}}^2\right]^{1/2}}
+ \sqrt{2 (1-b_{\mathrm{T}})} \right] \right] ~.~~~~~~
\end{eqnarray}

\noindent For the unfocused case where $b_{\mathrm{T}}=0$ (also nearly
applicable for the focused case if
$\epsilon_{\mathrm{ks}} B_{\mathrm{ks}}^2/B_{\mathrm{W}}^2 \geq 1$),
the maximum of this function occurs at $B=B_{\mathrm{W}}$, so 
$ M_{\mathrm{X}}=2 \sqrt{2} \sqrt{1-B_{\mathrm{W}}/B_{\mathrm{ks}} }$.
 For the focused case, when 
$\epsilon_{\mathrm{ks}} B_{\mathrm{ks}}^2/B_{\mathrm{W}}^2 \ll 1$
the maximum occurs when $b_{\mathrm{ks}} = b_{\mathrm{T}} = 1- \epsilon_{\mathrm{ks}}/ \sqrt{3}$, yielding
$ M_{\mathrm{X}}= 1.26 (B_{\mathrm{W}}/B_{\mathrm{ks}}) (1/ \sqrt{\epsilon_{\mathrm{ks}}})$.

The best MHD condition is achieved by making $M_{\mathrm{ks}}$ as small
as possible subject to the condition that $\beta_{\mathrm{ks}}$ always
remain less than unity. To fulfill MHD expectations and the
$\beta_{\mathrm{ks}}<1$ condition limits the allowable range for
$M_{\mathrm{ks}}$ in the unfocused case (or a focused case where $\Delta
\mu B_{\mathrm{ks}}/E_0 >1 $) to
\begin{eqnarray}
1>M_{\mathrm{ks}} = 0.95 \sqrt{1-B_{\mathrm{W}}/B_{\mathrm{T}} } 
\frac{\alpha_{\mathrm{c}} \beta_{\mathrm{c}}}{\sqrt{\sigma_{\mathrm{p}}} } g  ~.
\end{eqnarray}

\noindent For the focused case the permitted range of operation for
$M_{\mathrm{ks}}$ is
\begin{eqnarray}
1>M_{\mathrm{ks}} \geq  0.4 \frac{B_{\mathrm{W}}}{B_{\mathrm{ks}} } 
\frac{\alpha_{\mathrm{c}} \beta_{\mathrm{c}}}{\sqrt{\sigma_{\mathrm{p}}}  \epsilon_{\mathrm{ks}}^{1/2} } g ~.
\label{finmks}
\end{eqnarray}

\noindent  Thus the condition that the beta value of the kinetic
stabilizer beam be less than unity may limit the utility of obtaining
enhanced MHD stabilization by focusing the beam at $B=B_{\mathrm{T}}$
because the inequality in \eqref{finmks} could be violated for small
$\epsilon_{\mathrm{ks}}$.

Another possible limitation to the KSTM is that the upper ratio of
$B_0/B_{\mathrm{W}} = r^2_{\mathrm{W}}/r^2_0$ must satisfy the particle
orbit adiabaticity condition. The ion parallel velocity be significantly
larger than the curvature drift velocity, requiring that
\begin{eqnarray}
\sqrt{2E_0}\,\sigma_{\mathrm{ad}} > {2E_0\over\omega_{\mathrm{ci}}(z_{\mathrm{W}})} {d^2r(z_{\mathrm{W}})\over dz^2}
= \sigma_{\mathrm{p}}\,{E_0\over\omega_{\mathrm{ci}}(z_{\mathrm{W}})r_{\mathrm{W}}}\label{KS}~,
\end{eqnarray}

\noindent where $\sigma_{\mathrm{ad}} \sim 1$ is a constant associated
with the adiabaticity. We find
\begin{eqnarray}
B_{\mathrm{W}}({\mathrm{gauss}})\ge {0.2 E(\mathrm{keV})\over B_0(T)r^2_0(\mathrm{m})}
\left({\sigma_{\mathrm{p}}\over\sigma_{\mathrm{ad}}}\right)^2\left({m_{\mathrm{i}} (a)\over 2}\right).\label{constant}
\end{eqnarray}
For a $100\,$MW reactor where $B_0(T)r^2_0(m)\cong 1\,$Tm$^2$, the
adiabaticity constraint does not impose a serious restriction on the
choice of design parameters.

\section{Trapped Particle Instability\label{sectpm}}

Now let us consider the influence of the trapped particle mode to the
stability of the KSTM. The crucial aspect to establish stability
of the trapped particle mode is to have enough electrons that
communicate between the central-cell region and kinetic stabilizer
region. To determine if stabilization of the fast-growing trapped particle mode is possible,
 we estimate the fraction of electrons in the
kinetic stabilizer region that need to reflect back into the
central plasma  \cite{berksov}.

\begin{figure}[H]
\includegraphics[scale=.95]{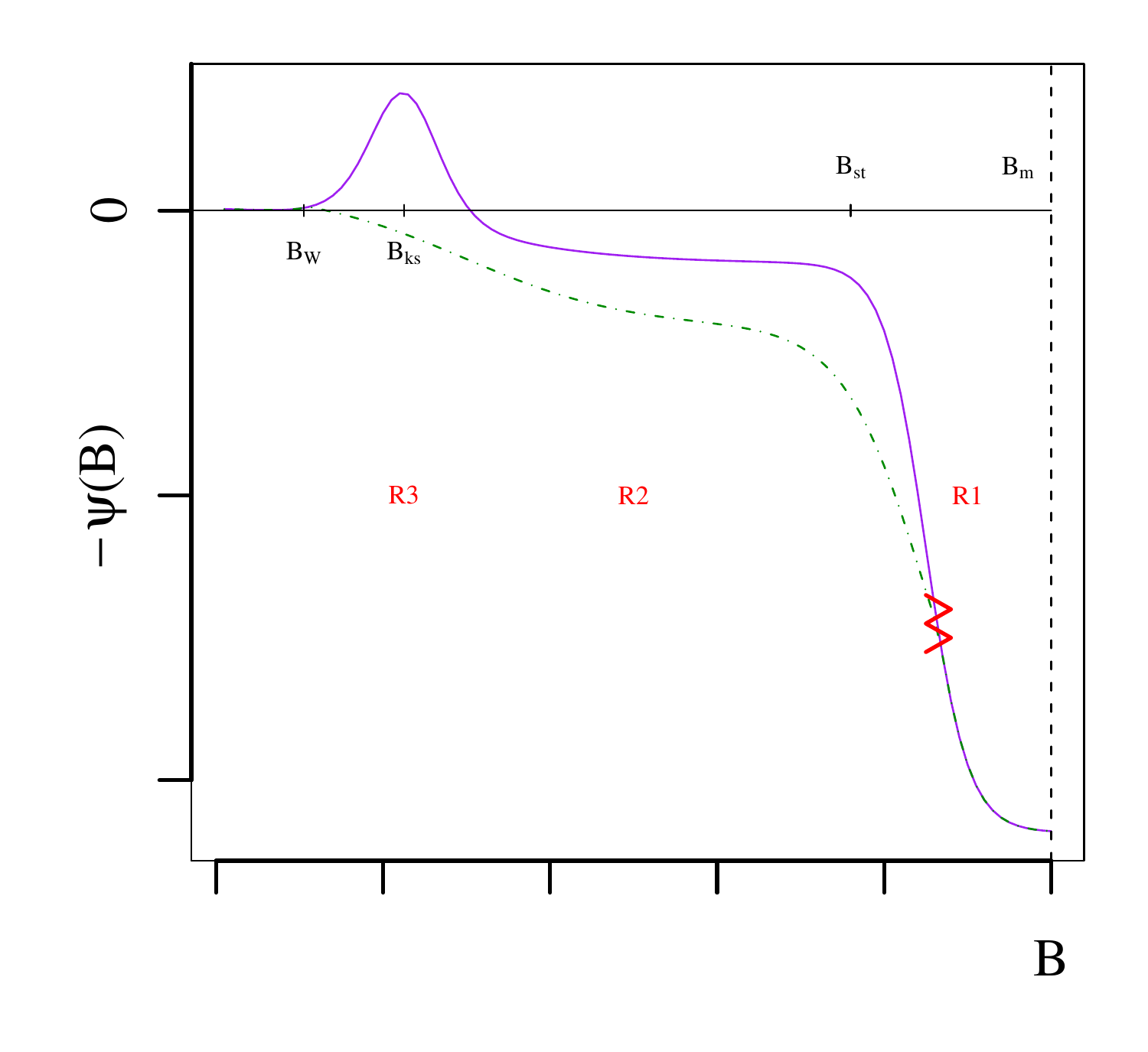}
\caption{Schematic diagram of the negative of the electrostatic potential energy felt by
the electrons for the focused (solid) and unfocused (dashed) cases. In
region R1, the ambipolar potential is determined by the electron
temperature of the tandem-mirror-confined plasma. Region~R2 is the
plateau region where the electron density and ambipolar potential are
nearly constant in space.  Region R3 is the kinetic stabilizer region
of large stabilizing concave curvature. }
\label{psi_ff5}
\end{figure}

To obtain a reasonable estimate requires we need to describe the structure of the
potential well in the region outside of the plugs. Immediately outside
of the MHD destabilizing plug (shown on the right-hand side of 
\fref{psi_ff5}) there exists an ambipolar potential energy of electrons,
$\Psi(B)=-|e|\Phi(B)$ (with the choice $\Phi_W=0$).  This ambipolar
potential forms in the region between the wall and the plugs, preventing
all but the most energetic electrons from escaping. The effective
potential acting on electrons is $U_{\mathrm{eff}}(B)=\Psi(B)+\mu B$.
Due to the opposite sign of their electric charge, the ambipolar
potential forces an immediate escape of any ions that reach the expander
region beyond the plug.

%FIG. 6

In the region labeled R1 in \fref{psi_ff5},
escaping ions experience an increase in kinetic energy,
$E=T_{\mathrm{ic}}+\kappa T_{\mathrm{e}0}$ that is several times the
central-cell electron temperature, with $\kappa\simeq 5$.  The ions in
this escaping beam are moving nearly parallel to the magnetic field.  We
assume that there is a rapid transition around a \emph{stand-off region}
where $B=B_{\mathrm{st}}$. At the stand-off region there is a transition
in the ambipolar potential; towards the central cell region (to the
right of $B_{\mathrm{st}}$ in \fref{psi_ff5}).  The ambipolar
potential energy varies on the scale of the central-cell electron
temperature.  Toward the kinetic stabilizer region (to the left of
$B_{\mathrm{st}}$ in \fref{psi_ff5}) the ambipolar potential tracks
with the significantly lower electron temperature of the expander. The
stand-off position may be determined at the point where the total stress
tensor of the effluent matches the pressure of the kinetic stabilizer
beam. However a proper theory for this intuitively described model of
the potential structure on either side of the abrupt stand-off region
still remains for future work.

%chunky
The characteristics of region R2 are determined by the kinetic
stabilizer beam.  As has been discussed in \sref{seckstmmodel}, in region R2,
between the KS region and the stand-off point, there exists a density
plateau where the ambipolar potential and ion density are nearly
spatially constant. This plateau region exists regardless of whether
the KSTM beam is focused or unfocused; however, the plateau region has
significantly higher density in the unfocused case.

We assume that the electrons in the kinetic stabilizer have a long mean
free path and can be described by a Maxwellian distribution with
temperature $T_{\mathrm{eks}}$. For simplicity, we take
$T_{\mathrm{eks}}\ll m_{\mathrm{i}} E_0$, although we stretch this
inequality to its limits of validity. The Maxwellian assumption can be
justified for the focused case where an ambipolar potential produces a
dominant fraction of trapped electrons (shown in \fref{psi_ff5} in region R3 as a
solid curve between $B_{\mathrm{W}} < B
<B_{\mathrm{ks}}$); the majority of the electrons are trapped and
confined long enough to relax by collisions to a Maxwellian distribution
\cite{prattdiss09}. For the unfocused case the determination of the mean
electron energy in the R2 and R3 regions is far more complex.  The electron distribution may
not be a Maxwellian distribution in this case.
For the unfocused case the ambipolar potential monotonically increases
to the wall values.

For the focused case, the schematic view of the ambipolar potential is
shown in the region R3 of \fref{psi_ff5}. This region is dominated
by a large population of trapped electrons that neutralize the incoming
ion beam but do not readily communicate with electrons in the central-cell.
 The bulk of the electrons in the kinetic stabilizer region are
insulated from the central-cell plasma by an ambipolar potential. The
case with maximum MHD stabilization is
particularly susceptible to the trapped particle instability \cite{prattdiss09}.

The unfocused case leads to moderately reduced MHD stabilization, but
has a far less restrictive condition for achieving trapped particle mode
stabilization. Because the electrostatic potential monotonically
increases in the direction from the expander toward the central-cell,
electrons with magnetic moment $\mu=0$ feel an inwardly-directed
electrostatic force. However, electrons with a finite magnetic moment
feel an outwardly-directed mirror force that by itself would produces a
magnetic trap what would prevent entrance to the central-cell.
Nevertheless we presume that if any electron from the distribution of
temperature $T_{\mathrm{eks}}$ penetrates through the stand-off point
$B=B_{\mathrm{st}}$, it will also penetrate into the central-cell
because of the large electrostatic forces directed toward the
central-cell for $B>B_{\mathrm{st}}$. For this to happen requires
$T_{\mathrm{eks}} B_{\mathrm{mx}}/B_{\mathrm{st}}<\kappa
T_{\mathrm{e}0}$, where $\kappa$ is a numerical constant of order unity.  However the spatial potential profile is an important
issue that may still prevent some electron penetration. For example, if
most of the large potential drop is very close to the plug, a large
fraction of the electrons would be reflected by the magnetic field and
reduce the trapped electron penetration into the central-cell. 
Nonetheless, we choose the most favorable case for satisfying the
trapped particle instability: the case where all electrons that reach
the abrupt stand-off section penetrate into the central region of the
plasma.

Ions moving toward the central-cell from the end wall do not have enough
energy to penetrate beyond the stand-off region since the large
ambipolar potential there repels them. These ions are blocked from
penetrating by both the repulsive quality of the magnetic mirror force
(accounted for in the distributions we have chosen) and
ultimately by the extremely strong electrostatic force in the transition
region that confines the hot central-cell
electrons. In contrast, electrons in the kinetic
stabilizer region, whose source is from a supply originating from the
end-walls, feel forces pushing them into the central-cell. Hence, there
are no confined ions sharing both the central-cell and kinetic
stabilizer region.  There are electrons that circulate both in the
central-cell and in the kinetic stabilizer region. These electrons
are a source for the charge uncovering term that is a stabilizing factor
for the trapped particle mode \cite{berksov}.

The MHD stability criterion that we use in this paper strictly is the
stability criterion for a low beta $m=1$ flute mode where the radially
perturbed displacement is constant everywhere along a cylindrical-like
column and we shall apply this criterion to the finite beta regime.
Consider a test function that pinches off the displacement so that there
is zero displacement in the regions of favorable curvature.  The trapped
particle mode arises  because the pinched-off potential removes the
stabilizing influence from the good curvature region.  In ideal MHD, the
pinching process excites bending energy that allows the continued
stabilization of the mode. However, kinetic processes enable the
excitation of an electrostatic perturbation that does not excite bending
energy. In addition, a low-beta flute mode is electrostatic, which
enables relatively efficient coupling to other electrostatic
perturbations.  In a tokamak such coupling leads to the
Kadomtsev-Pogutse trapped particle mode \cite{kptpm}, which allows a
curvature-driven instability to persist, but at a greatly reduced
growth-rate compared with the prediction from MHD theory. In a tokamak
this reduced growth-rate occurs due to the circulation of passing
particles through the entire plasma.  In a tandem mirror there is a
substantially lower fraction of connecting particles than in a tokamak
(where the fraction is near unity). Thus the relative growth-rate of the
trapped particle mode is larger relative to the trapped particle
growth-rate in a tokamak.  When the fraction of trapped particles is
sufficiently low and no additional stabilization mechanism is employed,
the growth-rate can even be as large as the MHD growth rate of the
central-cell and plugs. Mode stabilization of the trapped particle mode
in a tandem mirror relies on the property that there can be a different
fraction of connecting electrons to connecting ions. This difference
leads to a stabilizing effect, which is considered below.

The stabilization condition for the $m=1$ mode does not have the usual
finite Larmor radius stabilization \cite{rkr62}. However, a similar
stabilization mechanism arises due to \emph{charge uncovering}, the
difference between the number of electrons and ions that sample both the
central cell and the kinetic stabilizer regions. In ideal MHD, the
cross-field currents due to the lowest order $\vec{E} \times \vec{B}$
drift do not produce an electrical current due to cancellation of
electron and ion flow velocities. Because of finite Larmor radius
effects, the ion $\vec{E} \times \vec{B}$ drift differs from the
electron $\vec{E} \times \vec{B}$ drift. This leads to a current that
produces charge accumulation.  For trapped particle and for
electrostatic modes, a similar current emerges due to the difference in
electron and trapped ion particle populations.

We analyze the case where density and pressure profiles are
Gaussian and have the form $R(z,
\psi)=r(z)\exp{\left[-\psi/2\psi_0\right]}$ where $\psi = \psi(0) = B_0 r_0^2/2 = B(z) r^2(z)/2$.   We follow past
trapped particle mode studies in describing the electrons in the trapping region
(in this work the KS region) by a Maxwellian distribution
at a fixed temperature, $T_{\mathrm{eks}}$.  Neglecting the effect of rotation found in
their work, Berk and Lane\cite{berklane862} found the stability condition to be:
\begin{equation}
\left(\omega_{\mathrm{e}}^*\Delta Q\right)^2 > 4\gamma^2_{\mathrm{MHD}} (1+Q) ~,
\end{equation}

\noindent where $\omega^*_{\mathrm{e}}= T_{\mathrm{eks}}/|e| B_0 r^2_0$.  The square of the MHD growth rate $\gamma^2_{\mathrm{MHD}}$ is defined to be
\begin{equation}
\gamma^2_{\mathrm{MHD}} ={\displaystyle\int\limits_{-L}^L dz r^3_0(z)\displaystyle{d^2
r_0(z)\over dz^2}\left(p_{\perp}(z)+p_{||}(z)\right)\over\displaystyle\int\limits_{-L}^L dz r^4_0(z)\rho_{\mathrm{m}}(z)} ~.
\end{equation}

\noindent where $\rho_{\mathrm{m}}(z)$ is the mass density.  $Q$ is defined
\begin{eqnarray}
Q&=&{2\left(B_0 r^2_0\right)^2\displaystyle\sum\limits_j
\displaystyle\int\limits_{\mathrm{ks}}
\displaystyle{dz\over T_j} r^2_0(z) n_{\mathrm{ct}j}(z) 
e^2_j\over\displaystyle\int\limits_{-L}^L 
dz r^4_0(z)\rho_{\mathrm{m}}(z)} ~,
\end{eqnarray}

\noindent and
\begin{eqnarray} 
\omega_{\mathrm{e}}^*\Delta Q &=&{-2B_0 r^2_0\displaystyle\sum\limits_j
\displaystyle\int\limits_{\mathrm{ks}}\displaystyle{dz} r^2_0(z)
n_{\mathrm{ct}}(z) e_j
\over\displaystyle\int\limits_{-L}^L dz r^4_0(z)\rho_{\mathrm{m}}(z)} ~.
\end{eqnarray}

\noindent where $j$ denotes the species of particle with charge $e_j$.  
The density of particles of species $j$ at position $z$ that connect is
$n_{\mathrm{ct}j}(z)$,  \emph{i.e.} these particles
travel directly between, the kinetic stabilizer region and the central-cell when
a long mean-free-path electron limit is assumed.

In the kinetic stabilizer region the eigenmode excitation is assumed to
be negligibly small.  The ambipolar sheath prevents ions outside the
central region from returning to the central region, so that only
electrons connect to the central-cell. Thus in the expression for $Q$
and $\Delta Q$, only electrons contribute to the sums.  For this case
$\Delta Q=Q$, so that we find fulfillment of the trapped particle
stability criterion requires,
\begin{eqnarray}
Q>2\left({\gamma_{\mathrm{MHD}}\over\omega^*_{\mathrm{e}}}\right)^2\left(1+
\sqrt{1+\left({\omega^*_{\mathrm{e}}\over \gamma_{\rm
MHD}}\right)^2}\right)\xrightarrow[\text{$\left(\displaystyle{\omega^*_{\mathrm{e}}\over \gamma_{\rm
MHD}}\right)^2\ll 1$}]{}4\left({\gamma_{\mathrm{MHD}}\over\omega^*_{\mathrm{e}}}\right)^2.
\label{tpmstabnc}
\end{eqnarray}

\noindent If this
condition is not met, the growth-rate of the trapped particle mode is
approximately given by \cite{berklane862}
\begin{eqnarray}
\gamma_{\mathrm{tp}}\approx\gamma_{\mathrm{MHD}}/\sqrt{1+Q} .
\end{eqnarray} 
If $Q<1$ the intrinsic growth rate is purely MHD, as if  no kinetic
stabilizer were present.

For simplicity, as the stability condition for \eqref{tpmstabnc}, we
only use $Q > 4(\gamma_{\mathrm{MHD}}/\omega^*_{\mathrm{e}})^2$ of
\eqref{tpmstabnc}, a  \emph{necessary condition} for the stability
of the kinetic stabilizer to the trapped particle mode. In terms of
physically intuitive parameters, the necessary
stability condition is
\begin{eqnarray}
T_{\mathrm{eks}}\displaystyle\int\limits_{\mathrm{ks}}
dzr^2_0(z)n_{\mathrm{ct}}(z)>4\displaystyle\int\limits_0^{L_{\mathrm{c}}+L_{\mathrm{pl}}}dz\left(p_\perp(z)+p_\|
(z)\right)r^3_0(z){d^2 r_0(z)\over dz^2}\label{condition-} ~.
\end{eqnarray}

\noindent where $\mathrm{ks}$ refers to the $z$ integration over a single kinetic stabilizer region.

We investigate conditions that satisfy both MHD and trapped particle
stability. To clarify the stability boundary we consider the MHD marginality parameter
$M_{\mathrm{ks}}$ defined in \eqref{machine}.  For an MHD stable system
\begin{equation}
M_{\mathrm{ks}} =\frac{\displaystyle\int\limits_0^{L_{\mathrm{c}}+L_{\mathrm{pl}}} dz
\left(p_\perp+p_\|\right)\ r^3_0\frac{d^2 r_0}{dz^2}}{\displaystyle\int\limits_{\mathrm{ks}}
dz\left(p_\perp+p_\|\right)\ r^3_0\frac{d^2 r_0}{dz^2}}<1.\label{KSstab}
\end{equation}

\noindent It is convenient to express the trapped particle stability condition
relative to the MHD stabilization drive of the kinetic stabilizer as:
\begin{equation}
\frac{T_{\mathrm{eks}}\displaystyle\int\limits_{\mathrm{ks}} dzr^2_0(z)
n_{\mathrm{ct}}(z)}
{4\displaystyle\int\limits_0^{L_{\mathrm{c}}+L_{\mathrm{pl}}} dz\left(p_\perp(z)+p_\|(z)\right)
r^3_0\,\frac{d^2 r_0}{dz^2}}
\\
= \frac{ T_{\mathrm{eks}} \lambda_{\mathrm{k}} \sigma_{\mathrm{p}} }
{4 m_{\mathrm{i}} E_0 M_{\mathrm{ks}} }  \left<{n_{\mathrm{ct}}\over n}\right>
>1\label{relation}~.
\end{equation}

\noindent In this expression, $\left \langle n_{\mathrm{ct}}/ n\right \rangle$ is the fraction of electrons that have orbits in the kinetic stabilizer and also
penetrate into the central-cell.  This fraction is
\begin{eqnarray}
\left<{n_{\mathrm{ct}}\over n}\right> =
{\displaystyle\int\limits_{\mathrm{ks}}\displaystyle{dz\over B}
n_{\mathrm{ct}}(z)\over\displaystyle\int\limits_{\mathrm{ks}}\displaystyle{dz\over B}\
n(z)}~.\label{connfrac}
\end{eqnarray}

\noindent The numerical factor $\lambda_{\mathrm{k}}$ from the ratio in \eqref{relation} is
\begin{eqnarray}
\lambda_{\mathrm{k}}&=&{\displaystyle\int\limits_{\mathrm{ks}} dz\displaystyle\int\limits_0^1 
\displaystyle{dx\over(1-x)^{1/2}}{\left(\displaystyle{\Delta\mu B\over E_0}\right)\over
\left(x-{B\over B_{\mathrm{T}}}\right)^2+
\left(\displaystyle{\Delta\mu B\over E_0}\right)^2}\over\displaystyle\int\limits_{\mathrm{ks}} dz\displaystyle\int\limits^1_0 \displaystyle{dx\over(1-x)^{1/2}}(2-x){\left(\displaystyle{\Delta\mu
B\over E_0}\right)\over\left(x-\displaystyle{B\over B_{\mathrm{T}}}\right)^2+\left(\displaystyle{\Delta\mu B\over
E_0}\right)^2}}
\\
&\cong& {\displaystyle\int\limits_{\mathrm{ks}} dz\displaystyle{1\over\left(1-\displaystyle{B\over B_{\mathrm{T}}}\right)^{1/2}}\over\displaystyle\int\limits_{\mathrm{ks}} dz{2-\displaystyle{B\over B_{\mathrm{T}}}\over\left(1-\displaystyle{B\over B_{\mathrm{T}}}\right)^{1/2}}}.
\label{lamk}
\end{eqnarray}

%\begin{eqnarray}
%\lambda_{\mathrm{k}}&=&m_{\mathrm{i}} E_0 \psi_0 {
%{\displaystyle\int\limits_{\mathrm{ks}}\frac{dz}{B}n(z)}}\over{
%\int\limits_{\mathrm{ks}} dz p(z)  r^3_0\,\displaystyle{\frac{d^2 r_0}{dz^2}}}
%\\
%&=& \psi_0 {\displaystyle\int\limits_{\mathrm{ks}} dz\displaystyle\int\limits_0^1\displaystyle{dx\over(1-x)^{1/2}}{\left(\displaystyle{\Delta%\mu B\over E_0}\right)\over\left(x-\displaystyle{B\over B_{\mathrm{T}}}\right)^2+
%\left(\displaystyle{\Delta\mu B\over E_0}\right)^2}\over\displaystyle\int\limits_{\mathrm{ks}} 
%dz    r^3_0\,\displaystyle{\frac{d^2 r_0}{dz^2}} \displaystyle\int\limits^1_0 \displaystyle{dx\over(1-x)^{1/2}}(2-x){\left(\displaystyle{\Delta\mu
%B\over E_0}\right)\over\left(x-\displaystyle{B\over B_{\mathrm{T}}}\right)^2+\left(\displaystyle{\Delta\mu B\over E_0}\right)^2}}
%%\cong {\displaystyle\int\limits_{\mathrm{ks}} dz\displaystyle{1\over\left(1-\displaystyle{B\over B_{\mathrm{T}}}\right)^{1/2}}\over\displaystyle\int\limits_{\mathrm{ks}} dz{2-\displaystyle{B\over B_{\mathrm{T}}}\over\left(1-\displaystyle{B\over B_{\mathrm{T}}}\right)^{1/2}}}.
%\label{lamk}
%\end{eqnarray}

%fuzzy bookmark

\noindent The last approximate integral in \eqref{lamk} applies when the ion beams
satisfy 
\begin{eqnarray}\nonumber
\Delta\mu B_{\mathrm{ks}}/(E_0-B/B_{\mathrm{T}})\ll 1~.
\end{eqnarray}
\noindent For beams with 
$\Delta\mu B_{\mathrm{T}}/E_0\gg 1$, \emph{i.e.} unfocused beams,
we obtain $\lambda_{\mathrm{k}} \doteq 1/2$.  For well-focused beams, where 
$B_{\mathrm{T}} = B_{\mathrm{ks}}$ and $\Delta\mu B_{\mathrm{ks}}/E_0\ll 1$
we have $\lambda_{\mathrm{k}} \doteq 1$.

For a focused beam, the ion density build-up at $B\simeq B_{\mathrm{T}}$
allows for increased MHD stabilization of a factor of 1.5-3, as shown in \fref{foyay4}.  
  However the focusing creates an ambipolar potential that prevents
most of the electrons within the kinetic stabilizer from connecting to
the central-cell.   For this reason the focused case may have the best MHD
properties but is extremely susceptible to the trapped particle mode
because the factor $\left<n_{\mathrm{ct}}/n\right>$, the fraction of
trapped and connecting electrons, is exponentially small.

We therefore examine the unfocused case which will have a substantially
larger value of $\left<n_{\mathrm{ct}}/n\right>$; for this case it may be possible to
 satisfy the trapped particle stabilization criterion.
An unfocused beam results in $\lambda_{\mathrm{k}}\cong 0.5$ and the trapped
particle stability criterion is
\begin{equation}
\langle{{n_{\mathrm{ct}}\over n}} \rangle \ {1\over 8}{T_{\mathrm{eks}}\over m_{\mathrm{i}}  E_0} >
M_{\mathrm{ks}}.\label{trprst}
\end{equation}

\begin{figure}[H]
\includegraphics[scale=.9]{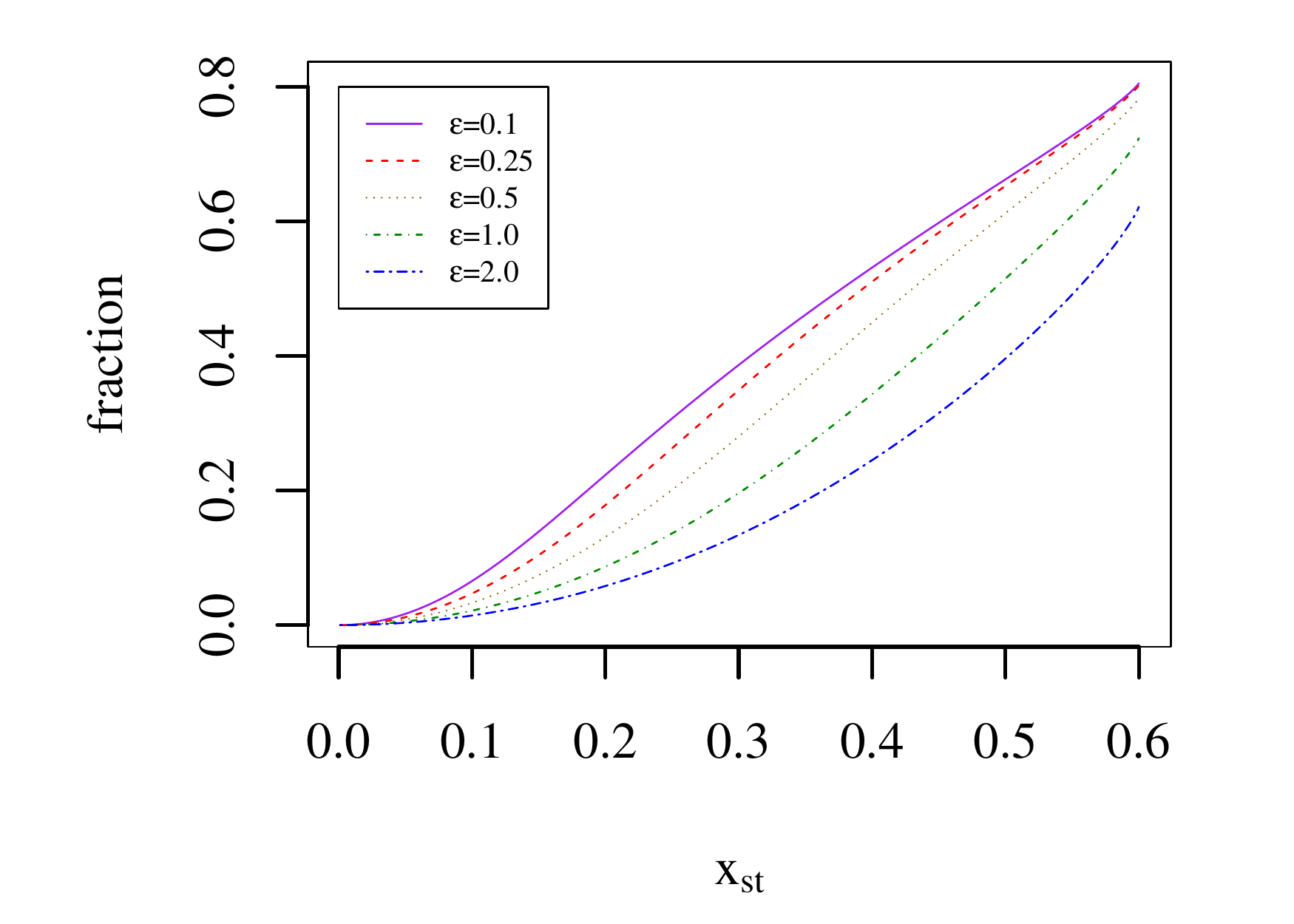}
\caption{The connecting fraction of electrons in the kinetic stabilizer region that
reach the central-cell. The relative radius of the kinetic stabilizer
entrance is $r_{\mathrm{ks}} /r_{\mathrm{W}} = 0.6$.  The various curves are for different
values of $\epsilon=\Delta\mu B_{\mathrm{ks}}/E_0$. The connecting fraction is plotted as a
function of the stand-off position $x_{\mathrm{ks}} = r_{\mathrm{ks}}/r_{\mathrm{W}}$.}
\label{connecting}
\end{figure}

To apply this stability condition, we have calculated the normalized
connecting particle electron density $n_{\mathrm{ct}}(z)/n(z)$
\begin{eqnarray}
{n_{\mathrm{ct}}(z)\over n(z)}=1-\sqrt{1-\displaystyle{B(z)\over B_{\mathrm{st}}}}
\exp{\left[-\left({\psi_{\mathrm{st}}-\psi(z)\over\displaystyle{B_{\mathrm{st}}\over B(z)}-1}\right) \right]},
\end{eqnarray}

\noindent where $\psi(z)= - |e| \phi(z)/T_{\mathrm{eks}}$.  The derivation of this
formula is given in the Appendix.  Recall
that the subscript st refers to the stand-off position.  The
connecting fraction of electrons $\langle n_{\mathrm{ct}} /n \rangle$, defined in \eqref{connfrac},
is evaluated numerically.  The results of this calculation are shown in  
\fref{connecting}.

%FIG. 7

Despite an increase value of $\left<n_{\mathrm{ct}}/n\right>$ compared
to the trapped case, it is still difficult to satisfy the stability
criterion given by \eqref{trprst} even when $M_{\mathrm{ks}}$ is
substantially less than unity. This is because all the factors on the
left-hand side of \eqref{trprst}
are relatively small.  For example, in \fref{connecting},
$\left<n_{\mathrm{ct}}/n\right>$ tends to be small, although it
can be
close to unity if the stand-off position can be made close to the inbound
entrance to the kinetic stabilizer region.  Furthermore, if one attempts to raise the
ratio $T_{\mathrm{eks}}/m_{\mathrm{i}}  E_0$ to an order unity quantity,
the emerging ambipolar potential may prevent ion penetration into
kinetic stabilizer region. We find in a calculation not
discussed here \cite{prattdiss09} that when a self-consistent
attempt to construct an ambipolar potential is made, raising
$T_{\mathrm{eks}}/m_{\mathrm{i}} E_0 \approx 0.3$ led to a
breakdown of the quasi-neutrality condition. Though higher
electron temperature solutions should be possible with the formation of
internal potential jumps on the order of a Debye length, it is likely
that the maximum ratio $T_{\mathrm{eks}}/m_{\mathrm{i}}  E_0$ that can give
penetration of the ion beam into the kinetic stabilizer will remain
below unity. 

To satisfy the trapped particle stability criterion given by
\eqref{trprst}, we need to make the MHD stability margin parameter
$M_{\mathrm{ks}}$ as small as possible; at the same time we must ensure
that the largest acceptable kinetic stabilizer beam power throughput is
less than the alpha particle power production of the central-cell.  It
then follows from \eqref{fpower} that $M_{\mathrm{ks}}$ lies in the
interval
\begin{equation}
1>M_{\mathrm{ks}}>70\frac{\lambda_{\mathrm{c}}}{r_0(\mathrm{m})}
\left( \frac{B_{\mathrm{W}} }{ B_0} E_0({\mathrm{keV}})\right)^{1/2}
\left(\frac{T_{\mathrm{c}}(100 ~{\mathrm{keV}}) }{\beta_{\mathrm{c}} } \right)
\left( \frac{3}{ B_0({\mathrm{T}})} \right)^2.
\label{eq69}
\end{equation}

We fold the trapped particle stability criteria of 
\eqref{trprst} in with the MHD power constraint given in 
\eqref{fpower}. For simultaneous stability to
the MHD and trapped particle modes, as well compatibility with
the power restrictions of the kinetic stabilizer beam, the kinetic stabilizer's window is
\begin{eqnarray}
{\mathrm{min}}\left[ 1, \left\langle \frac{n_{\mathrm{ct}}}{n} 
\right\rangle \sigma_{\mathrm{p}} \frac{T_{\mathrm{eks}}}{8 m_{\mathrm{i}} E_0} \right]
< M_{\mathrm{ks}} < 
\\ \nonumber 0.7
\frac{\lambda_{\mathrm{c}} }{r_0(\mathrm{m})} 
\left( \frac{B_{\mathrm{W}} }{B_0} E_0 (\mathrm{keV})\right)^{1/2}
\left( \frac{T_{\mathrm{c}} (100~\mathrm{keV}) } {\beta_{\mathrm{c}} } \right)
\left( \frac{3}{B_0 (\mathrm{T}) }\right)^2.~~~
\end{eqnarray}
\noindent The trapped particle is always more restrictive than the MHD
criterion.  We take $\lambda_{\mathrm{c}} =.175$, although design improvement from our nominal
parameter choice can reduce $\lambda_{\mathrm{c}}$.  Hence the compatibility criterion becomes
\begin{eqnarray} \nonumber
4.6\frac{(\lambda_{\mathrm{c}}/0.175)}{\sqrt{3} ~r_0 (\mathrm{m})   \sigma_{\mathrm{p}}}
\left( \frac{3 B_{\mathrm{W}}(\mathrm{g}) }{B_0 (\mathrm{T}) E_0(\mathrm{keV}) } \right)^{1/2}
\left( \frac{T_{\mathrm{c}} (100~\mathrm{kev}) } {\beta_{\mathrm{c}} } \right)
\\
\left( \frac{3}{B_0 (\mathrm{T}) }\right)^2
\left( \frac{m_{\mathrm{i}} E_0}{2 T_{\mathrm{eks}} }\right)
\left\langle \frac{n}{2.5 n_{\mathrm{ct}} }\right\rangle <1~.~~~
\label{lasteq}
\end{eqnarray}

\noindent We have arranged the left hand side of \eqref{lasteq}, so
that each of the bracketed terms have a nominal value of unity. We see
difficulty in fulfilling this condition. We need experimental designs that will allow the bracketed
terms to achieve smaller values. Much of the detailed physics factors
are buried in the factor $\lambda_{\mathrm{c}}$.  Perhaps novel ideas
can be developed to gain a large reduction factor in this parameter. The
nominal value of other factors, such as $\left(m_{\mathrm{i}}
E_0 / 2 T_{\mathrm{eks}} \right)$ and $ \left\langle n / 2.5
n_{\mathrm{ct}} \right\rangle $ have been selected to have as small a
value as deemed possible. Hence, the present theory for a collisionless
trapped particle mode indicates that there is a significant stability/power issue
for the stabilization of a symmetric mirror
machine with a kinetic stabilizer.

\section{Summary and Conclusions \label{secconc}} 
  
We have investigated the compatibility of the kinetic stabilizer to both
MHD stability and trapped particle stability.  With more beam input power, a smaller
value can be
obtained for the MHD stability parameter, $M_{\mathrm{ks}}$, which must
be less than unity to fulfill the MHD stability criterion. However, for a fusion device the power requirement for sustaining the
kinetic stabilizer beam must be substantially less than the fusion power
produced.

  In our analysis the nominal maximum beam power is taken
to be the fusion alpha power production, approximately 20 \% of the total
fusion power production.  Then the MHD and power constraints lead to an allowable range of values
for $M_{\mathrm{ks}}$, given by \eqref{finmks}. In addition, the
window of operation for simultaneous fulfillment of power requirements
and trapped particle stability leads to the relation given by 
\eqref{lasteq}. As it stands, the trapped particle instability
criterion, together with power constraints, would not be satisfied in a
burning plasma. Designs that improve on our choices for the of nominal
parameters is needed. Below, other caveats to this conclusion are
discussed.

A pertinent issue is how severe the trapped particle instability can be.
A systematic experimental study of this instability has yet to be
undertaken. In the kinetic stabilizer region, if the electrons are in
the short mean-free-path regime, the trapped particle instability growth
rate is likely to decrease.  Then it may be feasible to implement
feedback techniques to prevent or reduce the harmful effects of the
trapped particle instability. Further studies, especially experimental
studies, are needed to establish a data base to assess the implications
of exciting the trapped particle instability and determine whether the
harmful effects of this instability can be mitigated.

Our study also indicates the need to develop calculations that can
systematically calculate the ambipolar potential in the kinetic
stabilizer regime. The shape of the electric field in the case of
unfocused beam propagation is particularly challenging to evaluate,
because electrons in the kinetic stabilizer are in contact with the
wall. In one transit, electrons are unlikely to be described by a
Maxwellian distribution; it is necessary to verify how reasonable the
model used in this work is for describing the potential structure in the KS region.
A successful theory would enable the determination of where the
stand-off position would lie.
 
There is an additional concern regarding the trapped particle
stabilization criterion. If trapped particle stability is achieved by
improving the communication of electrons from the kinetic stabilizer
beam to the central cell, there may be a break down in the thermal
insulation of the hot electrons in the center of the machine.  Suppose a
value of $\langle n_{\mathrm{ct}}/n \rangle \sim 0.3$ is achieved, a
reasonable estimate of the fraction of connecting electrons necessary to
fulfill the trapped particle instability criterion. Then the current of
cold electron entering the central cell will be 30 \% of the kinetic
stabilizer beam current. These connecting electrons are accelerated to
the peak energy of the ambipolar potential and replace more energetic
electrons that leave the system. The energy lost per electron in the
exchange is comparable to the central-cell electron temperature. The
thermal loss rate would be $\langle n_{\mathrm{ct}}/n \rangle
P_{\mathrm{ks}} T_{\mathrm{ec}}/E_0$, where $T_{\mathrm{ec}}$ is the
electron temperature in the central cell and $P_{\mathrm{ks}}$ is the
power sustaining the kinetic stabilizer. Since $T_{\mathrm{ec}}$ will be
very large ($\geq 10^4$) the connection of the particles, lead to an
unacceptable power drain.

Another concern is that the local beta achieved by a focused kinetic
stabilizer beam may exceed unity. Such a plasma is formed by
injecting a beam narrowly distributed in magnetic moment so that the
beam will reflect back to the wall at designated target field position
$B=B_{\mathrm{ks}}$. At that position there is large favorable
field-line curvature with the magnetic field designed so that
$dr_{\mathrm{ks}}/dz=0$. Such a condition leads to a logarithmically
large, stabilizing MHD response. Mixing the response of a focused beam
with a completely unfocused beam may enable the best satisfaction of
both MHD and trapped particle instability. 
The increase in the MHD
response is logarithmic in the small parameter $\delta \equiv
\sqrt{(\Delta r \Delta \mu)/(r_{\mathrm{ks}} \mu_{\mathrm{T}})}$, 
where $r_{\mathrm{ks}}$ is the plasma field-line radius at $z=z_{\mathrm{ks}}$
and $\Delta r$ is the spread in the focusing position of the injected particle beam.
The local beta
increases as $\delta^{-1/2}$. Thus if significant MHD enhancement is
achieved, the local beta value at $B_{\mathrm{ks}}$ is likely to be
substantially larger than unity as has been discussed in \sref{secbetalim}.
 An issue then arises regarding whether the desired focusing can
be achieved.

%\vspace{2mm}
%\noindent {\bf{Acknowledgements}}:  
\bigskip\noindent\ignorespaces
    \section*{Acknowledgments}
 Interesting discussions with Dimitry Ryutov, Richard
Post, and Wendell Horton are gratefully acknowledged.

\appendix
\section{Calculation of the density of connecting electrons}

We wish to derive an expression for the density of connecting electrons,
$n_{\mathrm{ct}}(z)$. The electrons are assumed to have Maxwellian
distribution of the form,
\begin{eqnarray}
f\left(v^2\right) = \frac{1}{\sqrt{2\pi}} \exp \left[\psi(z)-\left({v^2_\perp\over 2} 
+ {v^2_\|\over 2}\right)\right],
\end{eqnarray}

\noindent where the normalization had been chosen so that the electron
kinetic energy and ion potential energy $\psi$ are in units of
$T_{\mathrm{eks}}/m_{\mathrm{e}}$ and that local density is 
\begin{eqnarray} 
n(z)=n_{\mathrm{W}}\displaystyle\int\limits^\infty_{-\infty} dv_\|\displaystyle\int\limits^\infty_0 dv_\perp v_\perp
f\left(v^2\right) = n_{\mathrm{W}}\exp{\left[\psi(z)\right]}.
\end{eqnarray}

\noindent The connecting particles are those that reach the stand-off position
$z=z_{\mathrm{st}}$ with a non-zero $v_{||}$.  It follows from energy
conservation and magnetic moment conservation that the connecting
particles satisfy the condition

\begin{eqnarray}
{v^2_\perp\over 2} + {v^2_\|\over 2}-\psi(z)\ge {v^2_\perp B_{\mathrm{st}}
\over 2B(z)}-\psi_{\mathrm{st}}.
\label{A3}
\end{eqnarray}

\noindent Thus
\begin{eqnarray}
v^2_{\perp} \leq {v^2_\|+2\left[\psi\left(z_{\mathrm{st}}\right)-\psi(z)\right]\over
\displaystyle{B_{\mathrm{st}}\over B(z)}-1} \equiv v^2_{\perp\, \mathrm{mx}}(v_\|) ~.
\label{A4}
\end{eqnarray}

\noindent The connecting density is then evaluated by integrating over
the electron Maxwellian distribution inside the kinetic stabilizer with
the inequality given by \eqref{A4} satisfied.
\begin{eqnarray}
n_{\mathrm{ct}}(z)&=&n_{\mathrm{W}}\displaystyle\int\limits^\infty_{-\infty}
{dv_\|\over\sqrt{2\pi}}\ \mathrm{exp}\left(-{v^2_\|\over 2}\right)\displaystyle\int\limits^{v^2_{\perp\, \mathrm{mx}}
\left(v_\|\right)}_0 {dv^2_\perp\over 2}\ \exp\left(-{v^2_\perp\over 2}+\psi(z)\right)~,~~~~~~~~~~\\
&=&n(z)\left(1-\displaystyle\int\limits^\infty_{-\infty} {dv_\|\over\sqrt{2\pi}}\ \exp
\left[-{v^2_\|\over 2}{\left[B_{\mathrm{st}}+\left(\psi_{\mathrm{st}}-\psi(z)\right) B(z)\right]\over B_{\mathrm{st}}-B(z)}\right]\right) ~,~~~~~~~~~~
\\
&=&n(z)\left(1-\left({B_{\mathrm{st}}-B(z)\over B_{\mathrm{st}}}\right)^{1/2}\ 
\exp\left[-{\left(\psi_{\mathrm{st}}-\psi(z)\right)B(z)\over B_{\mathrm{st}}-B(z)}\right]\right)\label{cnt} ~.
\end{eqnarray}

\bibliographystyle{unsrtnat}
\bibliography{pratt_tmp_kstm}

\end{document}